%% file: main.tex
\documentclass[journal]{IEEEtran}
\usepackage{amsmath,amssymb,amsfonts}
\usepackage{pifont}
\newcommand{\cmark}{\ding{51}}
\newcommand{\xmark}{\ding{55}}

\usepackage{algorithmicx}
\usepackage[linesnumbered,ruled,vlined]{algorithm2e}
\usepackage{amsthm}
\usepackage{graphicx}
\usepackage{textcomp}
\usepackage{paralist}
\usepackage{hyperref}
\usepackage{multirow}
\usepackage{url}
\usepackage{setspace}
\usepackage{tabu}
\usepackage{romannum}
\usepackage{booktabs}
\usepackage{enumitem}
\usepackage{arydshln}
\usepackage[normalem]{ulem}
\usepackage{threeparttable}
\usepackage{tcolorbox}
\usepackage{url}
\usepackage{xcolor}
\usepackage{soul}
\usepackage{cite}
\usepackage{amsmath,amssymb,amsfonts}
\usepackage{algpseudocode}
\usepackage{graphicx}
\usepackage{subcaption}
\usepackage{float}
\usepackage{multirow}
\usepackage{xspace}
\usepackage{booktabs}
\usepackage{url}
\usepackage[T1]{fontenc}

\usepackage{multirow}
\usepackage{pgfkeys}

\def\BibTeX{{\rm B\kern-.05em{\sc i\kern-.025em b}\kern-.08em
    T\kern-.1667em\lower.7ex\hbox{E}\kern-.125emX}}

\newcommand{\tool}{SOLAR\xspace} %HARVEST, SHURE, SPUR, SUPER
\newcommand{\toolnf}{SOLAR\textsubscript{no-filtering}}

\newcommand{\toolnn}{SOLAR\textsubscript{non-normalized}\xspace}
\newcommand{\toolon}{SOLAR\textsubscript{only-neg}\xspace}

\newcommand{\subtool}{BST\xspace}

\makeatletter
\newcommand{\thickhline}{%
	\noalign {\ifnum 0=`}\fi \hrule height 1pt
	\futurelet \reserved@a \@xhline
}
\newcolumntype{"}{@{\hskip\tabcolsep\vrule width 1pt\hskip\tabcolsep}}
\makeatother

\begin{document}
\title{Listening to Users' Voice: Automatic Summarization of Helpful App Reviews}
% Release Planning of Mobile Apps Based on 

\author{Cuiyun~Gao,~Yaoxian~Li,~Shuhan~Qi,~Yang~Liu,~Xuan~Wang,~Zibin~Zheng, and~Qing~Liao

\thanks{The first two authors contribute equally to the work.}

\thanks{Accepted for publication by the IEEE Transactions on Reliability.}}
% revised August 26, 2015.


\markboth{Journal of \LaTeX\ Class Files,~Vol.~14, No.~8, August~2015}%
{Shell \MakeLowercase{\textit{et al.}}: Bare Demo of IEEEtran.cls for IEEE Journals}



\maketitle

\begin{abstract}
App reviews are crowdsourcing knowledge of user experience with the apps, providing valuable
information 
% details
for app release planning, such as major bugs to fix and important features to add. 
There exist prior explorations on app review mining for release planning,
% Prior explorations have been conducted on mining app reviews for release planning, 
however, most of the studies strongly rely on pre-defined classes or manually-annotated reviews. Also, the new review characteristic, \textit{i.e.}, the number of users who rated the review as helpful, which can help capture important reviews, has not been considered previously.

In the paper, we propose a novel framework, named \tool, aiming at accurately summarizing helpful user reviews to developers. The framework mainly contains three modules: The review helpfulness prediction module, topic-sentiment modeling module, and multi-factor ranking module. The review helpfulness prediction module assesses the helpfulness of reviews, \textit{i.e.}, whether the review is useful for developers. The topic-sentiment modeling module groups the topics of the helpful reviews and also predicts the associated sentiment, and the multi-factor ranking module aims at prioritizing semantically representative reviews for each topic as the review summary. Experiments on five popular apps indicate that \tool is effective for review summarization and promising for facilitating app release planning.
\end{abstract}

\begin{IEEEkeywords}
user reviews, review helpfulness, topic modeling, topic sentiment, review summarization
\end{IEEEkeywords}

\IEEEpeerreviewmaketitle

\input{sections/intro}
\input{sections/background}
\input{sections/approach}

\input{sections/setup}
\input{sections/experiment}

\input{sections/discussion}
\input{sections/literature}

\section{Conclusion}\label{sec:conclusion}

% A successful app relies on positive user experience and well-designed functions. 
To maintain high-quality apps, developers often take a lot of effort to extract key information from large amounts of scribbled user reviews.
In the work, we propose a novel framework, named \tool, focusing on automatically summarizing helpful user reviews for developers. 
\tool filters no-informative reviews based on a trained review helpfulness prediction model, and groups topics jointly with corresponding sentiments by the topic-sentiment summarization module. We also propose a multi-factor ranking module for prioritizing reviews for each topic. Extensive experiments verify the effectiveness of our proposed framework. In the future, we will conduct evaluation using app reviews across platforms and deploy \tool in industry.

% \appendices
% % you can choose not to have a title for an appendix
% % if you want by leaving the argument blank
% \section{}

% \section*{Acknowledgment}

\ifCLASSOPTIONcaptionsoff
  \newpage
\fi



\bibliographystyle{IEEEtran}
\bibliography{sigproc}


% \end{thebibliography}


% \end{IEEEbiography}


%\vfill


\end{document}

%% file: sections/intro.tex
\section{Introduction}
% user review mining is important
% user review mining 的challenges
% our proposed framework
% contributions
The quality of mobile apps directly influences the user experience and concerns the benefits gained by developers. With more apps continuing to spring up, app owners face more challenges in providing good service to users and standing out from competitors. User reviews are valuable information from users and reflect instant user experience with apps, \textit{e.g.}, major bugs encountered by users and missing app features. Summarizing the useful information in user reviews can help developers pay attention to important user concerns and thus facilitate release planning of the apps.

Online reviews not only enhance user awareness, but also serve as a reliable source of information about the quality of the app or the service of interest. Recently, user review mining has been extensively studied by both academic and industrial communities, on prioritizing app reviews~\cite{chen2014ar,DBLP:conf/issre/GaoWHZZL15,DBLP:journals/jss/PalombaVBOPPL18}, classifying reviews into different categories~\cite{DBLP:conf/icse/VillarroelBROP16,di2016would,DBLP:conf/re/MaalejN15}, predicting the features favored/disliked by users~\cite{DBLP:conf/kbse/GuK15,DBLP:conf/re/GuzmanM14}, or identifying emerging app issues~\cite{gao2019diver,gao2018online}. Most of the studies, however, strongly rely on pre-defined classes or manually-annotated reviews, which may require huge manual labor. For example, Chen \textit{et al.}~\cite{chen2014ar} observed that manually annotating 2,000 user reviews as informative or non-informative could cost 7.4 hours. Popular apps, such as Facebook and WeChat, may receive tens of thousands of reviews each day~\cite{appannie}. Thus, an automatic and effective approach is necessary for summarizing user reviews.

Automatically summarizing user reviews is challenging. First, user reviews are generally short in length and contain massive noisy words, \textit{e.g.}, misspelled words, and abbreviations, so the context information is limited. Second, user reviews are mostly non-informative. According to~\cite{chen2014ar}, only 30\% of the reviews provide informative user opinions for app updates, which increases the difficulty of extracting useful content from reviews. 
% \yx{which increases the challenge of obtaining useful content from reviews.}
Third, reviews contain multiple and various topics for different apps, and the pre-defined granularities are difficult to cover all the topics of the apps.
% to be consistent for all the studied apps.
% \yx{Third, reviews comprise several and distinct topics for different apps, and the described granularities are challenging to be consistent for all the investigated apps.}
For example, Noei \textit{et al.}~\cite{noei2019too} identified 23 common topics, such as searching and web browsing; while Di Sorbo \textit{et al.}~\cite{di2016would} summarized 12 topic clusters, including pricing and resources, etc., which are more general compared to Noei \textit{et al.}'s definition~\cite{noei2019too}. 
% Moreover, not all the topics demand developers' deep inspection, and accurate prioritization of the topics can be time-saving but still challenging. 
Moreover, not all the topics require in-depth inspection by developers, and appropriate prioritization of the topics can be time-saving but still challenging.
% For example, during prioritizing the extracted topics, ratings of user reviews are a commonly-used index, but the ratings may not be aligned with the review texts
For example, ratings of user reviews are a commonly-used index for prioritizing the extracted topics, but the ratings may not be aligned with the review texts
~\cite{DBLP:conf/hicss/MudambiSZ14}. 
%这里感觉转折有点生硬
% \textcolor{red}{We are motivated by the challenges and aim at proposing an end-to-end approach to adaptively capture and prioritize review topics in this work.}

% Specifically, 
To mitigate the above challenges, we design a novel framework, named \tool, an abbreviation for \textbf{S}ummarizati\textbf{O}n of he\textbf{L}pful \textbf{A}pp \textbf{R}eviews.
% \textbf{S}ummarization of hel\textbf{P}ful \textbf{U}ser \textbf{R}eviews. 
% au\textbf{TO}mati\textbf{C}ally \textbf{S}ummar\textbf{I}zi\textbf{N}ing user reviews. 
In Chen \textit{et al.}~\cite{chen2014ar}'s work, the informative reviews are extracted by training on manually-annotated reviews, which is rather labor-intensive and time-consuming. To alleviate the effort in filtering reviews according to the informativeness, we employ a new review characteristic, \textit{i.e.}, the number of users who rated the review as helpful, referred to as ``helpfulness number'' for convenience in the paper. The helpfulness number of each review indicates the volume of users who consider the delivered information is useful for them. In general, the reviews described in more detail or with real messages tend to be rated as more helpful~\cite{DBLP:journals/dss/Ngo-YeS14}, thus with the helpfulness number considered, the informative reviews could be captured, which constitutes the first process, \textit{i.e.}, review helpfulness prediction. Then, to mitigate the short-length nature of user reviews, we employ a Biterm Topic Model (BTM)~\cite{DBLP:conf/www/YanGLC13} for clustering topics, where BTM is specially designed for modeling topics of the short text corpus. For accurately estimating the sentiment associated with each topic, we adopt the
% propose a novel 
topic model approach
% , named 
\subtool~\cite{DBLP:journals/corr/abs-2008-09976}, 
% by 
which jointly models
% jointly modeling 
topics with sentiments. The topic modeling process is the second process, \textit{i.e.}, topic-sentiment summarization process. Finally, we propose a novel review ranking mechanism by involving multiple factors, including the semantic representativeness of the extracted topics and corresponding estimated sentiment, etc.

To validate the effectiveness of the proposed review summarization approach \tool, we conduct extensive experiments on five apps with a total of 11,659 reviews. Experimental results indicate that \tool achieves superior performance over the baseline approaches, increasing the precision and recall scores by at least 10.41\% and 12.75\%, respectively. 
% We also implement our approach as a tool\footnote{\yun{Add url.}} for readers to interact with.

The contributions of our paper are summarized as follows:

\begin{itemize}
\item We propose a novel framework for automatically and accurately summarizing user reviews for facilitating release planning of mobile apps.

\item We propose to predict the informativeness of reviews based on a new review characteristic, \textit{i.e.}, the helpfulness number, and no manual labor is required. Novel multi-factor review ranking approaches are also put forward for more accurate review summarization.

\item Experiments on real-world applications verify the effectiveness of the proposed framework. Our code and dataset are publicly available at \url{https://github.com/monsterLee599/SOLAR}.

\end{itemize}

\textbf{Paper structure.}
The remainder of this paper is organized as follows.
Section~\ref{sec:background} describes the motivation and background of our work.
Section~\ref{sec:approach} presents our proposed framework.
We introduce the experimental setup in Section~\ref{sec:setup}, and elaborate on the dataset, baseline models, and comparison results in Section~\ref{sec:exper}. Section~\ref{sec:discussion} provides some discussion about the proposed framework. Section~\ref{sec:literature} illustrates the related work. We conclude and mention future work in Section~\ref{sec:conclusion}.

%% file: sections/background.tex
\section{Background and Motivation}\label{sec:background}

\subsection{User Review and the Informativeness}
\begin{figure}[ht]
    \centering
    \begin{subfigure}[b]{0.45\textwidth}
        \includegraphics[width=\textwidth]{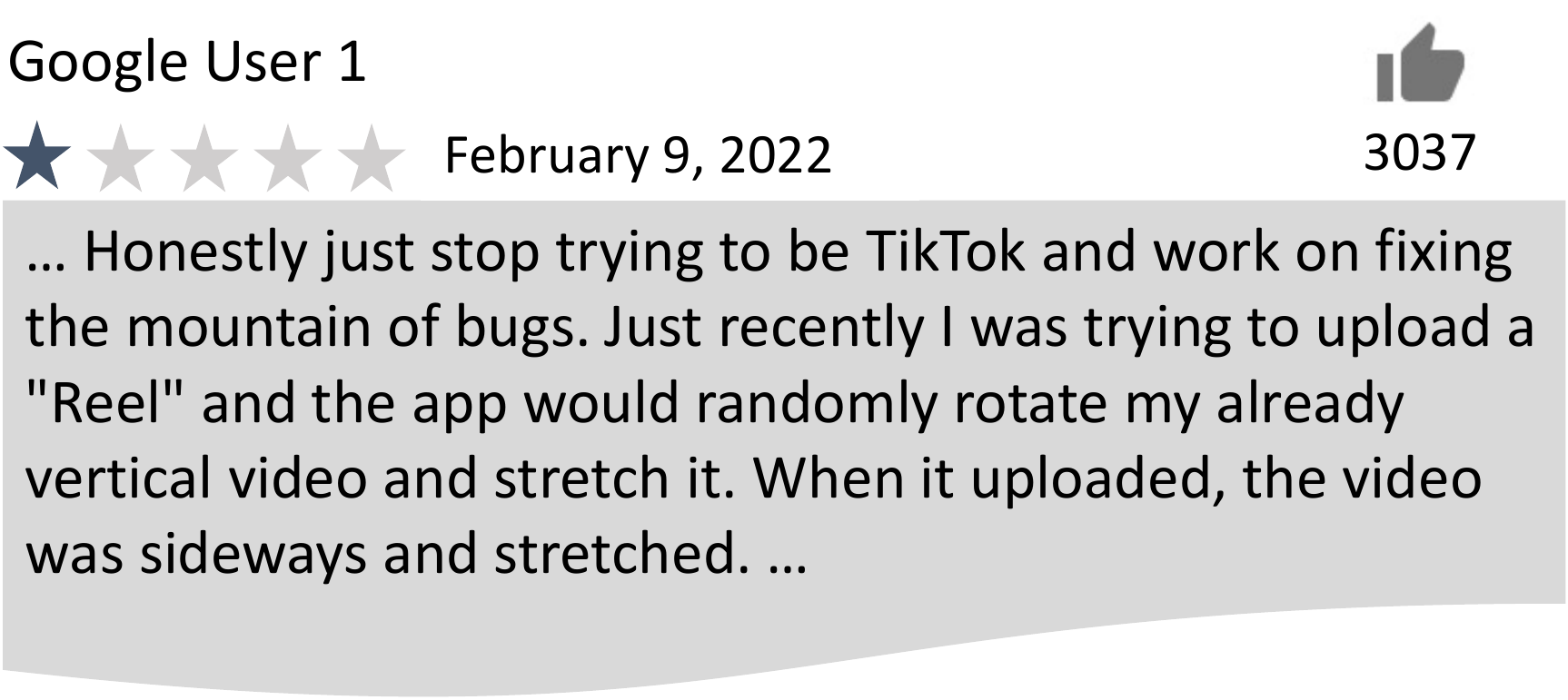}
        \caption{Example 1.}
        \label{fig:example_1}
      \end{subfigure}
      \hfill
      \begin{subfigure}[b]{0.45\textwidth}
        \includegraphics[width=\textwidth]{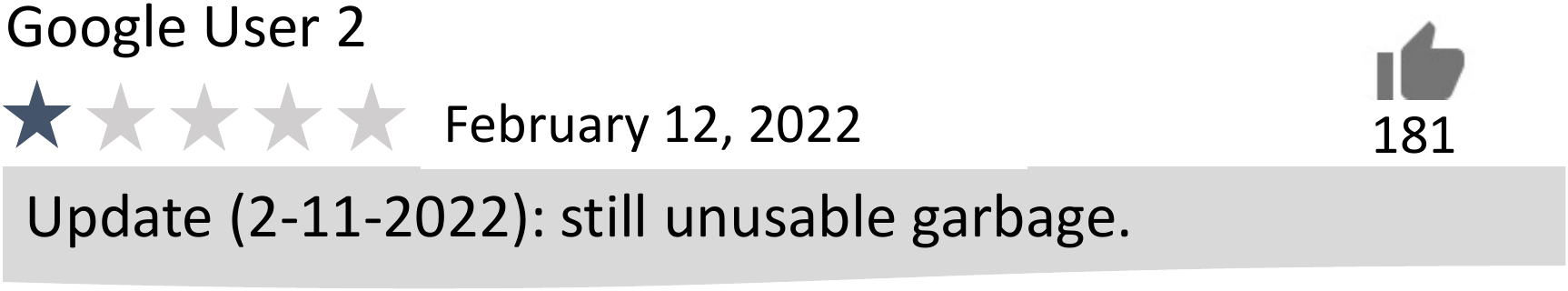}
        \caption{Example 2.}
        \label{fig:example_2}
      \end{subfigure}
    \caption{Illustration of the examples of user reviews.}
	\label{fig:examples}
\end{figure}

User reviews are an essential channel between app users and the developers, generally containing attributes such as user names, post dates, review texts, and ratings. Two examples of reviews for the Android Instagram app are illustrated in Figure~\ref{fig:examples}\footnote{The two examples were obtained on the same day from Google Play Store.}. Recently, Google Play releases a new characteristic, \textit{i.e. }, the number of users who rated the reviews as helpful, as shown at the top-right corner of each review. We refer to the new attribute as ``helpfulness number'' throughout the paper. We can see that the first piece of review, as depicted in Figure~\ref{fig:examples} (1), has a greatly larger helpfulness number than the second piece of review, possibly because review 1 provides more detailed and clearer feedback, \textit{e.g.}, about the ``\textit{reel}'' feature;
while review 2 only complains about the new update and does not detail the app issue. Thus, the attribute reveals the usefulness of the reviews to other readers, and can be considered as an index of the review's informativeness.
% To our best knowledge, no prior studies have employed such attribute for facilitating app review mining.

\subsection{Topic Modeling}
Topic modeling is a type of statistical model for uncovering the topics that occur in a collection of documents. One of the most popular topic modeling approaches is \textbf{Latent Dirichlet allocation (LDA)}~\cite{DBLP:conf/nips/BleiNJ01}.
% , with its graph model illustrated in Figure~\ref{fig:graph_models} (a). 
LDA assumes that each document is a mixture of topics, where a topic is a probabilistic distribution over words.
% As shown in Figure~\ref{fig:graph_models} (a), 
LDA models each document $r$ as a mixture of latent topics $\theta_r\in \mathbb{R}^{K}$ following a multinomial distribution, where $K$ is the number of topics. Each latent topic is described as a multinomial distribution $\phi\in \mathbb{R}^V$ over the vocabulary, where $V$ indicates the total number of unique words (\textit{i.e.}, vocabulary). 

% The whole modeling process can be formulated into the following form:
% \vspace{0.2cm}
% For each document $r$:
% \begin{itemize}
% \item Draw a topic mixture $\theta_r\sim Dirichlet(\alpha)$, where $\alpha$ is a hyper parameter.
% \item For each word $w_{n,r}$ in $r$:
%     \begin{itemize}
%         \item Draw a topic $z_{n,r}\sim Multinomial(\theta_r)$,
%         \item Draw a word $w_{n,r}\sim Multinomial(\phi_{z_{n,r}})$.
%     \end{itemize}
% \end{itemize}
% \vspace{0.2cm}

Although LDA has been proven successful in modeling formal and well-edited documents, such as news reports~\cite{DBLP:conf/nips/BleiGJT03} and scientific articles~\cite{DBLP:conf/uai/Rosen-ZviGSS04}, its performance will be inevitably compromised when processing short and ill-formed texts, such as app reviews and Twitter messages~\cite{DBLP:conf/www/YanGLC13}.

\textbf{Biterm Topic Model}~\cite{DBLP:conf/www/YanGLC13} is specifically designed for modeling topics in short texts. Different from LDA, which captures the document-level word co-occurrence patterns, BTM directly models the word co-occurrence patterns in the whole corpus.
% The graphical model is depicted in Figure~\ref{fig:graph_models} (b). 
The outputs of both LDA and BTM are two matrices: (1) Document-topic matrix $\Theta\in \mathbb{R}^{R\times V}$, where $R$ denotes the number of reviews; and (2) Topic-word matrix $\Phi \in \mathbb{R}^{K\times V}$.

\textbf{Joint Sentiment/Topic Model (JST)}~\cite{DBLP:conf/cikm/LinH09} can detect the topic sentiment besides modeling topics. Unlike other machine learning approaches for sentiment classification, JST is unsupervised. Different from LDA and BTM, JST assumes that topics are associated with sentiment labels and words are associated with sentiment labels and topics. JST also produces two matrices but with three dimensions: (1) Document-sentiment-topic matrix $\Theta\in \mathbb{R}^{R\times S\times K}$, where $S$ denotes the number of sentiment labels (\textit{e.g.}, $S=3$ indicate that the sentiment labels include positive, neutral, and negative); (2) Sentiment-topic-word matrix $\Phi\in \mathbb{R}^{S\times K\times V}$. 

% A graphical model of JST is represented in Figure~\ref{fig:graph_models} (c). 

% The formal definition of the modeling process of JST is as follows:

% \vspace{0.2cm}
% For each document $r$:
% \begin{itemize}
% \item Draw a topic mixture $\phi_{s,k}\sim Dirichlet(\gamma)$.
% \item Draw a topic sentiment distribution $\theta_{r, s}\sim Dirichlet(\alpha)$.
% \item For each word $w_{n,r}$ in document $r$:
%     \begin{itemize}
%         \item Draw a sentiment label $s_{n,r}\sim \pi_{r}$,
%         \item Draw a topic $z_{n,r}\sim Multinomial(\theta_{n,r})$,
%         \item Draw a word $w_{n,r}$ from the distribution over words defined by the topic $z_{n,r}$ and sentiment label $s_{n,r}$.
%     \end{itemize}
% \end{itemize}

% \begin{figure}[ht]
%     \centering
%     \begin{subfigure}[b]{0.13\textwidth}
%         \includegraphics[width=\textwidth]{figs//lda_graph.pdf}
%         \caption{LDA.}
%         \label{fig:lda}
%       \end{subfigure}
%       \hfill
%       \begin{subfigure}[b]{0.13\textwidth}
%         \includegraphics[width=\textwidth]{figs//BTM_graph.pdf}
%         \caption{BTM.}
%         \label{fig:btm}
%       \end{subfigure}
%       \hfill
%       \begin{subfigure}[b]{0.18\textwidth}
%         \includegraphics[width=\textwidth]{figs//jst_graph.pdf}
%         \caption{JST.}
%         \label{fig:jst}
%       \end{subfigure}
%     \caption{Graph illustration of three popular topic modeling approaches, where $\phi$ and $\theta$ denote the topic word distribution and document topic distribution, respectively. }
% 	\label{fig:graph_models}
% \end{figure}

%% file: sections/approach.tex
\section{Methodology}\label{sec:approach}

\begin{figure}[t]
    \centering
    \includegraphics[width=0.5\textwidth]{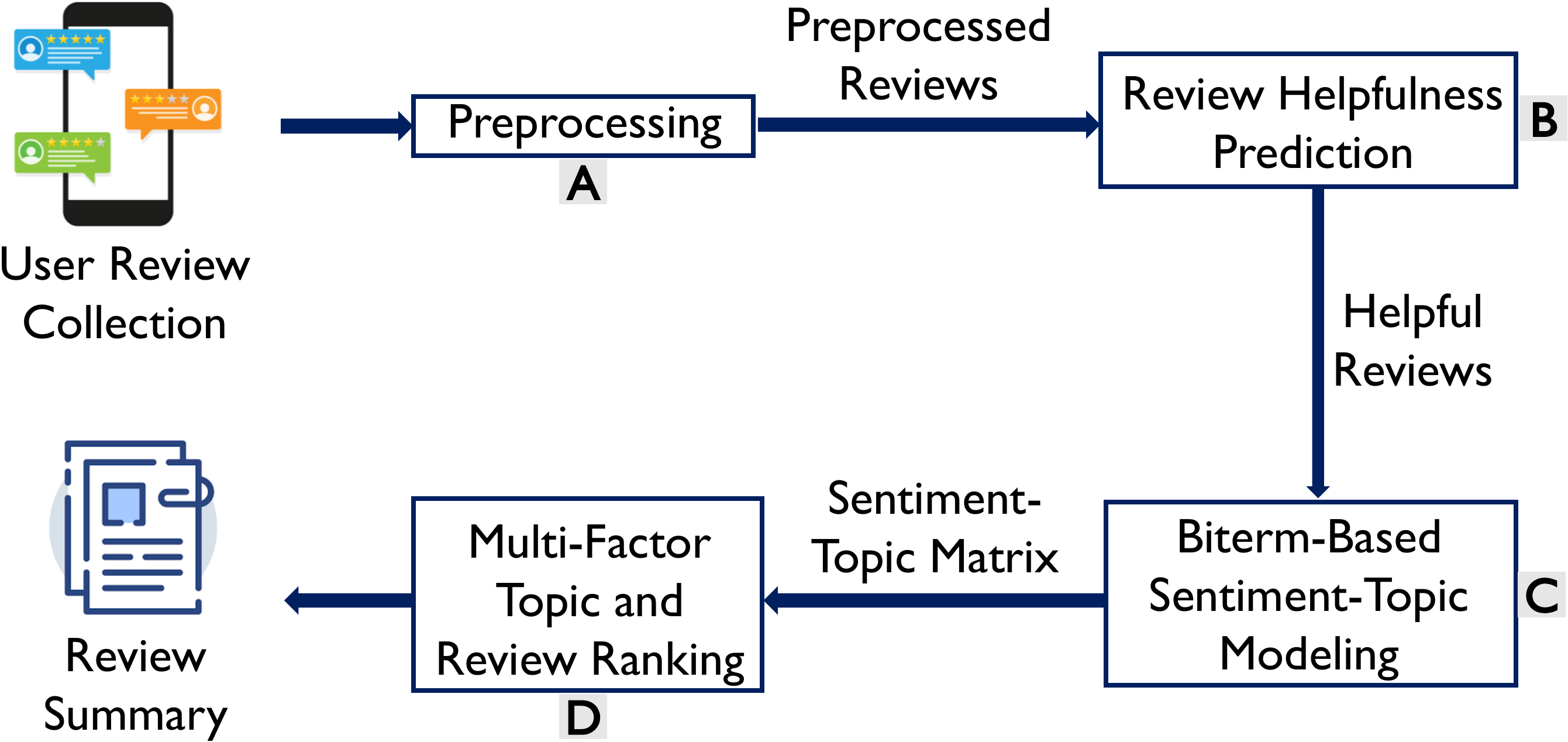}
    \caption{Overall architecture of the proposed framework \tool.}
	\label{fig:framework}
\end{figure}

In this section, we present an overview of the proposed framework \tool and then elaborate on each process of \tool. Figure~\ref{fig:framework} presents the overall architecture of the proposed framework, which consists of four major steps. The first step preprocesses the raw user review data into a well-structured format to facilitate subsequent processes. In the second step, the helpfulness of each review instance is estimated and the reviews predicted as ``helpful'' are delivered to the next step. The third step jointly models the topics and associated sentiment for the helpful reviews. The last step prioritizes (i) topics, and (ii) reviews instances for each topic based on multiple factors including semantic representativeness and sentiment. The prioritized reviews are regarded as the summary of reviews and will be provided to developers for managing app releases.

\subsection{Preprocessing}
We first remove emotions and convert all the reviews into their lowercase. We than adopt rule-based methods in~\cite{DBLP:conf/issre/ManGLJ16} to rectify repetitive words (\textit{e.g.}, ``\textit{very very good}'' to ``\textit{very good}''). Finally, we lemmatize each word into the root form following the lemmatization method described in~\cite{DBLP:conf/kbse/VuNPN15}, (\textit{e.g.}, ``\textit{was}'' to ``\textit{be}'').

\subsection{Review Helpfulness Prediction}

We extract 20 linguistic features which can potentially impact the helpfulness of review instances and differentiate helpful reviews from unhelpful ones. Tables~\ref{tab:feature} summarizes the set of 20 features which are grouped along 5 dimensions: stylistics, readability, lexicon, sentiment, and content.

\begin{table*}[t]
    \centering
    \caption{Summary of extracted features for review helpfulness prediction.}
    \label{tab:feature}
    \begin{tabular}{l|l|p{9cm}}
    \hline
        \textbf{Dimension} & \textbf{Feature Name} & \textbf{Description} \\
        \hline
        \hline
        \multirow{6}{*}{\textbf{Stylistics}} & Review-length & Number of words in the review \\ 
        & Sentence-length & Number of sentences in the review \\
        & Avg-sentence-length & Average number of words every sentence \\
        & 1-char-word-num & Percentage of the words consisting of only one character \\
        & 2-char-word-num & Percentage of the words consisting of 2 characters \\
        & \textgreater2-char-word-num & Percentage of the words consisting of \textgreater2 characters \\
        \hline
        \hline
        \multirow{5}{*}{\textbf{Readability}} 
        & Difficult-word-num & Number of the words difficult for understanding \\
        & Flesch & A metric for quantifying the readability of a text~\cite{farr1951simplification,DBLP:journals/eswa/Krishnamoorthy15a} \\
        & Dale-chall & A metric for quantifying the readability of a text~\cite{chall1995readability,Singh:article}\\
        & Misspelling-word-num & Number of the misspelled words\\
        \hline
        \hline
        \multirow{5}{*}{\textbf{Lexicon}} 
        & Noun-num & Number of nouns in the reviews \\
        & Verb-num & Number of verbs in the reviews\\
        & Adjective-num & Number of adjectives in the reviews\\
        & Subjective-num & Number of subjective words \\
        & Lexicon-diversity & Percentage of unique words to the total words \\
        \hline
        \hline
        \multirow{3}{*}{\textbf{Sentiment}} & Polarity & The polarity of the review, \textit{i.e.}, negative, positive, or neutral \\
        & Sentiment-word-num & Percentage of opinion words\\
        & Rating-extremity & Rating difference with the average app rating \\
        \hline
        \hline
        \multirow{3}{*}{\textbf{Content}} & Quality-related-word-num & Number of quality-related words\\
        & Uncertainty-degree & Number of the words indicating uncertainty\\
        & Unigram-tf-idf & The $tf\text{-}idf$ weight for each appeared word in the review \\
        \hline
    \end{tabular}
\end{table*}

\textbf{Stylistics Dimension} refers to the stylistic features including the numbers of words and sentences from word level, sentence level, and review level. The length information can influence the completeness of the information conveyed by reviews~\cite{DBLP:conf/issre/GaoWHZZL15,DBLP:journals/dss/SieringMR18,DBLP:conf/www/LuTNP10}. We use six features to quantify the stylistics dimension---namely, \textit{review-length}, \textit{sentence-length}, \textit{avg-sentence-length}, \textit{1-char-word-num}, \textit{2-char-word-num}, and \textit{\textgreater2-char-word-num}. All the features are calculated by counting words where the \textit{review-length} is from review level, \textit{sentence-length} and \textit{avg-sentence-length} are from sentence level, and the other features are from character level.

\textbf{Readability Dimension} refers to features that measure the readability of the user review. Readability, in general, is measured based on the syllables per word, the length of sentences, and the ratio of difficult words---it can estimate how many years of education are required for textual understanding~\cite{fan2018chaff}. To quantify the readability of the review text, we use the three readability measures proposed by previous work---namely \textit{difficult-word-num}, \textit{flesch}~\cite{farr1951simplification}, and \textit{dale-chall}~\cite{chall1995readability}.

\textit{Difficult words} are defined as those with more than two syllables, which do not including proper nouns, familiar jargon or compound words, and \textit{difficult words} do not contain common suffixes (\textit{e.g.}, ``-es'') as a syllable~\cite{gunning1952technique}. We denote the number of words, syllables, difficult words,  and sentences as \textit{Words}, \textit{Syllables}, \textit{Difficult Words}, and \textit{Sentences}, respectively. Based on the above definitions, the empirical formulas for calculating \textit{flesch}~\cite{farr1951simplification} and \textit{dale-chall}~\cite{chall1995readability} are shown as below:

\begin{equation}\label{equ:flesch}
    flesch = 206.835-1.015\frac{Words}{Sentences}-84.6\frac{Syllables}{Words}
\end{equation}
    
\begin{equation}\label{equ:dale}
    dale\text{-}chall = 0.16\frac{Difficult\;Words}{Sentences} + 0.05\frac{Words}{Sentences}
\end{equation}

\noindent where the constants in above formulas are from \cite{farr1951simplification,chall1995readability}.

We also consider the number of misspelling words as one index for readability, denoted as \textit{misspelling-word-num}. We define the misspelling words as those that are not found in Enchant English dictionary~\cite{Singh:article}.

\textbf{Lexicon Dimension} refers to the features that are related to the word lexicons. Four features are involved to quantify the lexicon dimension---namely, \textit{noun-num}, \textit{verb-num}, \textit{adjective-num}, \textit{subjective-num}, and \textit{lexicon-diversity}. We conduct part-of-speech tagging for each review text and count the respective numbers for nouns, verbs, and adjectives. The number of subjective words is counted based on the released subjective word list in~\cite{DBLP:journals/eswa/Krishnamoorthy15a}. The \textit{lexicon-diversity} is the ratio of the number of unique words in a review text to the \textit{review-length}.

\textbf{Sentiment Dimension} refers to the features reflecting user opinions. We consider three features for measuring the sentiment dimension---namely, \textit{polarity}, \textit{sentiment-word-num}, and \textit{rating-extremity}. The \textit{polarity} of a review text indicates whether the expressed opinion is positive, negative, or neutral. We measure the \textit{polarity} of the review text by computing the total positive score minus the total negative score of the review text, as illustrated in Equ.~(\ref{equ:polarity}). The positive score and negative score are computed as the numbers of positive words (denoted as \textit{Positive Words}) and negative words (\textit{i.e.}, \textit{Negative Words}), respectively, where the negative words and positive words are determined based on the SentiWordNet database~\cite{DBLP:conf/lrec/Esuli006}.

\textit{Sentiment-word-num} is calculated as the ratio of the number of sentiment words to the total words. We also consider \textit{rating-extremity} to measure how the review rating diverges from the average user rating, with the formula shown in Equ.~(\ref{equ:diverge}).

\begin{equation}\label{equ:polarity}
    polarity = \frac{Positive\; Words - Negative\; Words}{Words}
\end{equation}

% \begin{equation}\label{equ:sentiment}
%     sentiment\text{-}word\text{-}num = \frac{Positive\; Words + Negative\; Words}{Words}
% \end{equation}

\begin{equation}\label{equ:diverge}
    rating\text{-}extremity = ||Rating - Avg.\; Rating||
\end{equation}

\textbf{Content Dimension} refers to features that capture the textual characteristics based on text mining techniques. Prior work showed that text mining techniques can help with informative review extraction~\cite{chen2014ar}. Thus, we expect that analyzing the textual content of a review text can help distinguish \textit{helpful} and \textit{unhelpful} reviews. To quantify the content dimension, three features are included---namely, \textit{quality-related-word-num}, \textit{uncertainty-degree}, and \textit{unigram-tf-idf}.

Based on functionalities and features, users make assessments about actual app quality. We rely on the dictionary of General Inquirer~\cite{stone1963computer} to compute \textit{Quality-related-word-num}. The General Inquirer is a well-established framework for content analysis. The advantage of adopting the dictionary-based approach is the validation of the dictionary as well as the resulting standardized classifications~\cite{kelly1975computer}. The quality-related words are determined by their similarity distances with the word ``\textit{quality}'' in the General Inquirer dictionary. Examples of quality-related words are illustrated in Table~\ref{tab:example}. Similarly, we determine the review uncertainty score by considering the ``\textit{if}''-related words in the General Inquirer dictionary, which generally denotes feelings of uncertainty, doubt and vagueness~\cite{DBLP:journals/dss/SieringMR18} (exemplar words are ``\textit{almost}'' and ``\textit{may}'').

\textit{Unigram tf-idf} is a common technique for information retrieval and text mining, reflecting how important a word $w$ is to a review text $r$ in the collection $R$. The \textit{Unigram tf-idf} is calculated based on term frequency, \textit{i.e.}, \textit{tf}, and inverse document frequency, denoted as \textit{idf}.

\begin{equation}\label{equ:tf-idf}
tf\text{-}idf(w,r,R) = tf(w,r)\cdot idf(w,R),
\end{equation}

\noindent where $tf(w,r)=\log(1+freq(w,r))$ and $idf(w,R)=\log(\frac{|R|}{count\{r\in D:w\in r\}})$.

\textbf{Classification.} We characterize a review instance using the 20 features we extracted. The features are adopted to train a model to predict the helpfulness of a posted review. In this study, we use SVM (Support Vector Machine)~\cite{DBLP:journals/kais/WuKQGYMMNLYZSHS08} as the default classifier to construct the model.

\begin{table}[ht]
    \centering
    \caption{Review examples that express app quality or exhibit uncertainty. The respective indicating words are highlighted with wavy-underlined fonts.}
    \label{tab:example}
    \begin{tabular}{|l|p{5cm}|}
    \hline
    & \textbf{Example} \\
    \hline
    \hline
    \multirow{6}{*}{\textbf{Quality-Related}} & It is so \uwave{good} and \uwave{addictive} with \uwave{fantastic} graphics and so many games to choose from.\\
    \cline{2-2}
    &The effect is the app does super cool but sometimes isn't \uwave{smooth} and it gets annoying.\\
    \hline
    \hline
    \multirow{8}{*}{\textbf{Uncertainty-Related}} & I would really love this app \uwave{if} there was less popups and the daily quests were more realistic.\\
    \cline{2-2}
    & \uwave{If} you want players to feel like a purchase will help them stay in the game, then \uwave{perhaps} you should allow a few wins enough to keep them playing.\\
    \hline
    \end{tabular}
\end{table}

\subsection{Biterm-Based Sentiment-Topic Modeling}\label{sec:bbstm}

\begin{figure}[ht!]
    \centering
    \begin{subfigure}[b]{0.2\textwidth}
        \includegraphics[width=\textwidth]{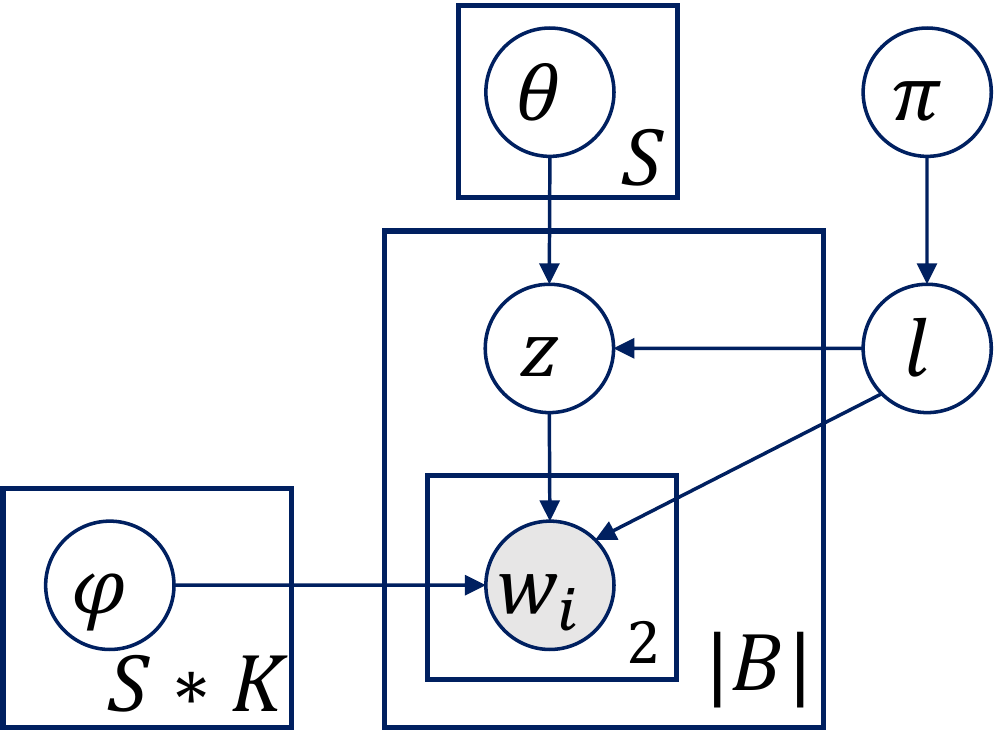}
        \caption{BST.}
        \label{fig:bst}
      \end{subfigure}
    \begin{subfigure}[b]{0.25\textwidth}
        \includegraphics[width=\textwidth]{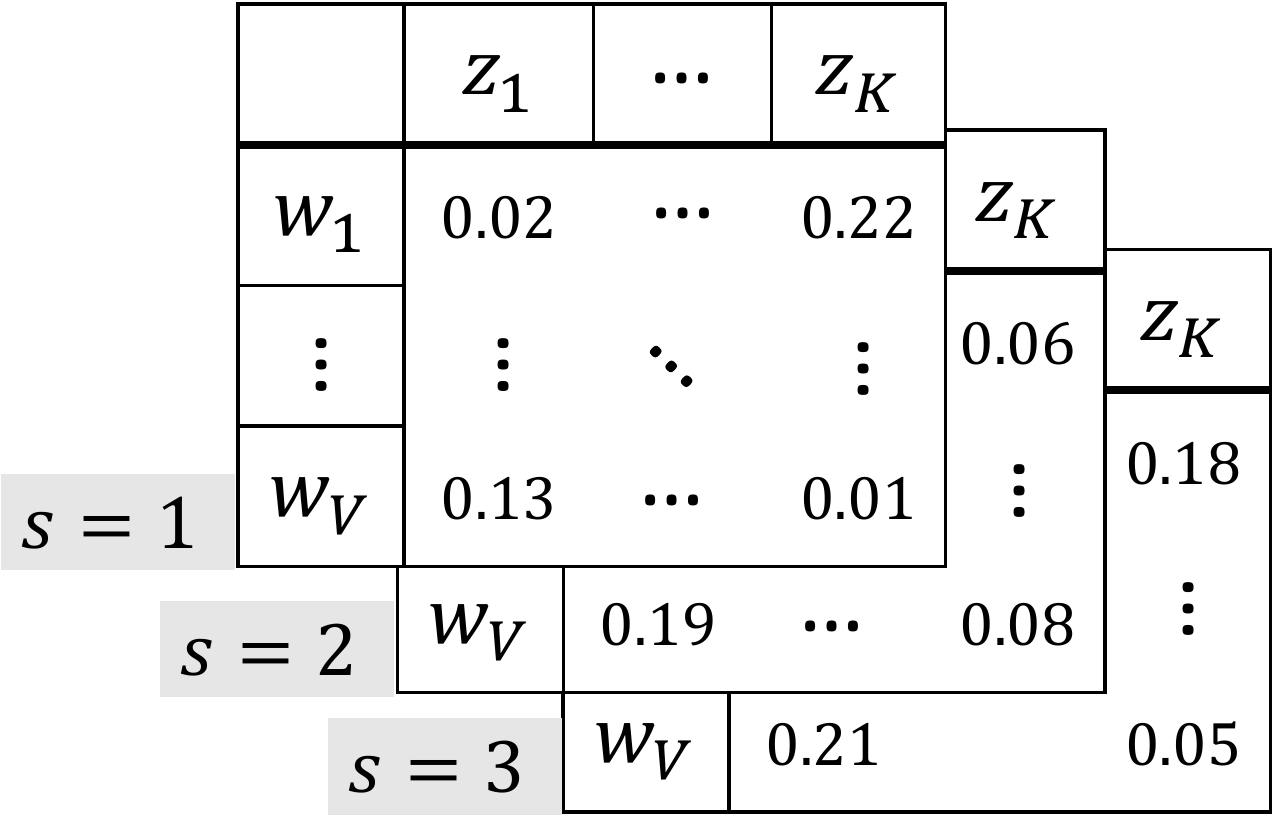}
        \caption{Example of the 3-D matrix $\Phi$.}
        \label{fig:matrix}
      \end{subfigure}
    \caption{Graph illustration of biterm-based sentiment-topic modeling, and example of the output sentiment-topic-word matrix $\Phi\in \mathbb{R}^{S\times K\times V}$.}
	\label{fig:bst-graph}
\end{figure}

Based on the prediction results of review helpfulness, we expect the reviews classified as ``\textit{helpful}'' to be informative for developers and employ them for the subsequent processes. In this section, we introduce the unsupervised model, named BST~\cite{DBLP:journals/corr/abs-2008-09976}, for jointly modeling topics and sentiment of app reviews. Figure~\ref{fig:bst-graph} (a) depicts the graph illustration of BST. The detailed modeling process can be referred to the work~\cite{DBLP:journals/corr/abs-2008-09976}.

An example of the output sentiment-topic-word matrix $\Phi$ is shown in Figure~\ref{fig:bst-graph} (b), where the number of sentiment labels is three and different sentiment labels $s=1$, $s=2$, $s=3$ indicate positive, neutral, and negative, respectively. For each biterm $b$, BST models its topic distribution over the vocabulary and sentiment distribution over the three sentiment polarities. The sentiment-topic distribution of each review $r$ can be calculated as:

\begin{equation}
    \begin{split}
    P(z, s|r) & = \sum_b P(z,s|b)\cdot P(b|r),\\
    P(z,s|b) & = \frac{P(z,s)P(w_i|z,s)P(w_j|z,s)}{\sum_{z,s}P(z,s)P(w_i|z,s)P(w_j|z,s)}, \\
    P(b|r) &= \frac{n_r(b)}{\sum_b n_r(b)},
    \end{split}
\end{equation}

\noindent where $n_r(b)$ is the frequency of the biterm $b$ in the review $r$. Similarly, we can infer the sentiment distribution of each review. We denote the computed topic distribution of each review $r$ as $P(z|r,s)=\{P(z_1|r,s), P(z_2|r,s), ..., P(z_k|r,s), ...\}$ where $k$ means the $k$-th topic given a sentiment label $s$ and $\sum_k P(z_k|r,s) = 1$. The sentiment distribution of each review is indicated as $P(s|r,z)=\{P(s_1|r,z), P(s_2|r,z), P(s_3|r,z)\}$ given a topic $z$, where $s_1$, $s_2$ and $s_3$ denote negative polarity, neutral polarity, and positive polarity, respectively.

\subsection{Multi-Factor Topic and Review Ranking}
\label{sec:ranking}
The multi-factor ranking step aims at prioritizing semantically representative reviews for each topic while ensuring the usefulness of the reviews for developers. We employ two procedures to rank reviews, \textit{i.e.}, topic ranking, and then review ranking.
The prioritization scores of topics are utilized for review ranking. \tool finally outputs the prioritized reviews for each topic.

\subsubsection{Topic ranking}

We prioritize the topics $z$ based on the corresponding features $\mathcal{F}^z$, including mainly four aspects: topic proportion, topic sentiment, average rating, and freshness. The total score for each topic is calculated as follows.

\begin{equation}
    Score_z = \sum_{f \in \mathcal{F}^z} \omega_f f,
\end{equation}
\noindent where $f \in \mathcal{F}^z$ is the grading aspect of each topic, $\omega_f$ is the weight of the computed score for aspect $f$, and $\sum_{f \in \mathcal{F}^z} \omega_f = 1$.

\textbf{Topic Proportion.}
Topics covering more reviews generally indicate that the topics have received more attention in the recent period; thus, the topics tend to be more important. We define the proportion for topic $z$ as below:
\begin{equation}
    f^{z}_{Volume} = \frac{\sum_{r \in \mathcal{R}} \sum_s P(z,s|r)}{|\mathcal{R}|max_{r \in \mathcal{R}}(P(z,s|r))},
\end{equation}
where $\mathcal{R}$ is the collection of reviews and $s$ indicates the sentiment label.

\textbf{Topic Sentiment.}
Generally, the reviews predicted as \textit{negative} tend to be more important for app developers than the positive reviews for app updating. Based on the output of Section \ref{sec:bbstm}, we calculate the sentiment score of a topic as follows. 
\begin{equation}
    f^{z}_{Sentiment} = \frac{\sum_{r \in \mathcal{R}} P(s_{1}|r,z)}{|\mathcal{R}|max_{r \in \mathcal{R}}(P(s_{1}|r,z))},
\end{equation}
\noindent where $s_1$ indicates the negative polarity.

\textbf{Average rating.}
The rating reflects the overall attitude of users towards the app. Topics with poorer ratings should be paid more attention by developers. We calculate the average rating of the topic $z$ as below.
\begin{equation}
    f^{z}_{Avg.Rating} =\frac{\sum_{r \in \mathcal{R}} Rating_r}{|\mathcal{R}|max_{r \in \mathcal{R}}(Rating_r)},
\end{equation}
where $Rating_r$ denotes user rating of the review $r$.

\textbf{Freshness.}
Latest reviews can embody users' newest opinions about the app, while the problems reflected in early-posted reviews tend to be relatively less important. We calculate the freshness feature of one topic as below.
\begin{equation}
    f^{z}_{Freshness} = \frac{\sum_{r \in \mathcal{R}} Timestamp_r}{|\mathcal{R}|max_{r \in \mathcal{R}}(Timestamp_r)},
\end{equation}
where $Timestamp_r$ indicates the post time of the review $r$.

\vspace{0.2cm}

\subsubsection{Review ranking}

The review ranking process considers a set of features $\mathcal{F}^r$ besides the prioritization scores of the topics. Other features include user rating, freshness, sentiment polarity, review length, and topic score. The overall score of one review $r$ is calculated as below.

\begin{equation}
    Score_r = \sum_{f \in \mathcal{F}^r} \omega_f f.
\end{equation}
where $f$ is the scoring feature of each review, $\omega_f$ is the weight defined for the feature $f$, and $\sum_{f \in \mathcal{F}^r} \omega_f = 1$.

\textbf{Rating.} User rating could directly express users' experience during the app usage. Poor user ratings generally indicate that the users are discontent with the app usage, and the reviews may describe the problems they encountered or unsatisfied with app functionalities. We normalize the user rating for each review as the rating score $f^{r}_{Rating}$.

% Rating can reflect the user's attitude towards the app, so we use rating as a scoring feature. In addition, we normalize rating for fair comparison as follow,
\begin{equation}
    f^{r}_{Rating} =\frac{ Rating_r}{\max(Rating)}.
\end{equation}
\textbf{Freshness.} We also consider the post time of the reviews. Reviews uploaded more recently could be more important to developers for app release.
% The latest reviews reflect issues that are more important to developers, so we use the review time as a feature. In addition, in order to give a fair score, we normalize the timestamp as follow,
\begin{equation}
    f^{r}_{Freshness} = \frac{Timestamp_r}{\max_{r \in \mathcal{R}}(Timestamp_r)}.
\end{equation}

\textbf{Sentiment Polarity.} The reviews predicted with negative sentiment polarity are more important for app release than the reviews with positive sentiment. Based on the sentiment-topic-word matrix in Section~\ref{sec:bbstm}, we calculate the sentiment score of each review as below.

% Through the sentiment-topic-word matrix in Sec.\ref{sec:bbstm}, we could calculate the different sentiment tendencies of review in its topic. And as discussed in topic ranking, the sentiment would influence the importance of review.
\begin{align}
    &f^{r}_{Negative} = P(s_{1}|r,z_k),&\\
    &f^{r}_{Neutral} = P(s_{2}|r,z_k),&\\
    &f^{r}_{Positive} = P(s_{3}|r,z_k).
\end{align}

\textbf{Review length.}
Reviews with longer lengths tend to deliver more detailed information about the user experience, and thus could be more useful for developers. The feature score is computed as below.

\begin{equation}
    f^{r}_{length} = -\log (Length_{r}),
\end{equation}
where $Length_{r}$ indicates the number of words in the review $r$.

\textbf{Topic.} The prioritization scores of the topics are also incorporated since the reviews are more related to the topics with higher ranking scores $Score_{z}$ would be more representative of the collected reviews. 

\begin{equation}
    f^{r}_{Topic} = \sum_{z} P(z|r)*Score_{z}.
\end{equation}

%% file: sections/setup.tex
\section{Experimental Setup}\label{sec:setup}

\subsection{Dataset}
\begin{figure}[t]
    \centering
        \includegraphics[width=0.4\textwidth]{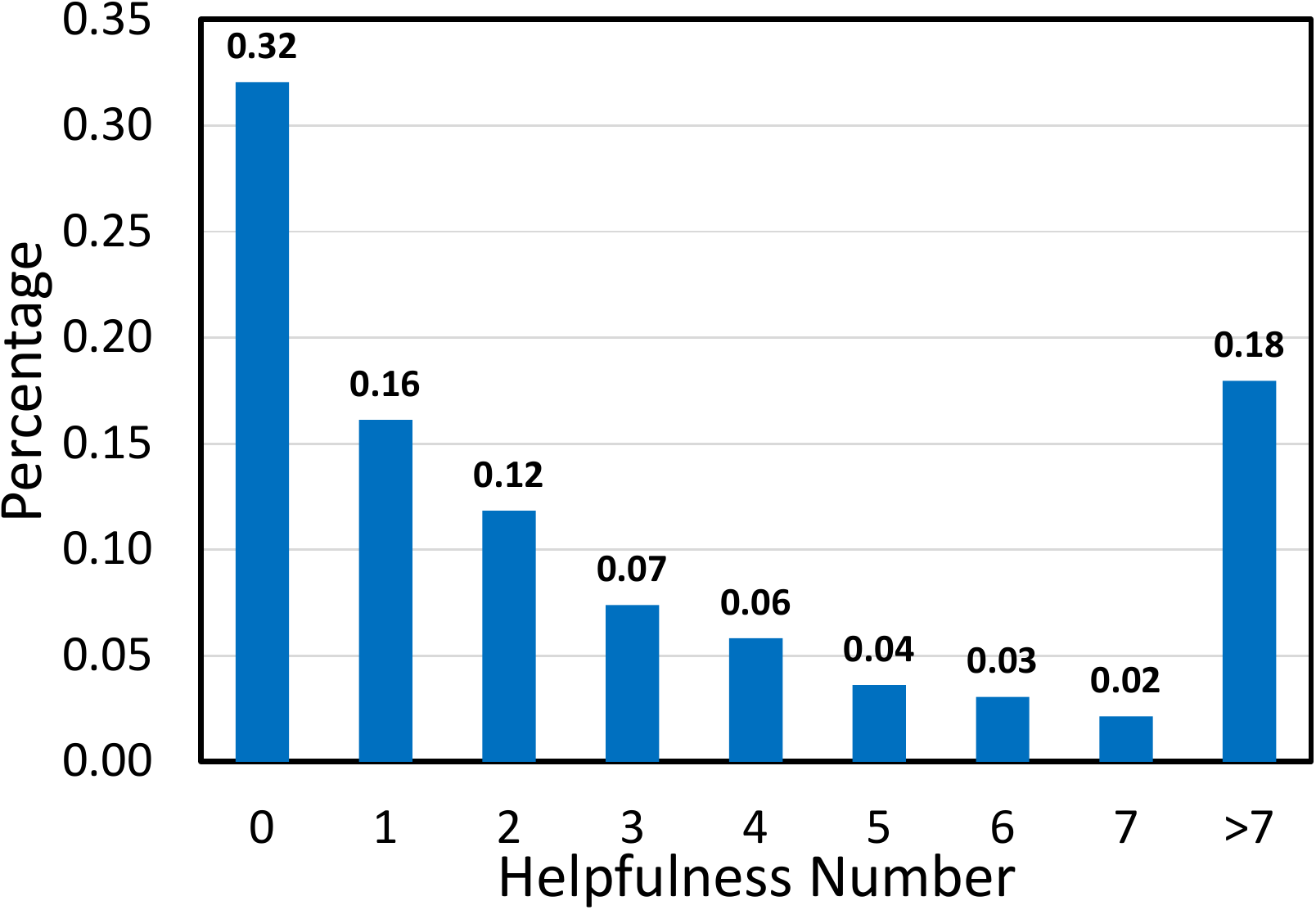}
    \caption{Distribution of helpfulness numbers for the collected reviews.}
	\label{fig:distribution}
\end{figure}

\textbf{Dataset for helpfulness Prediction.} 
We crawled the apps ranked at the top 200 during August 2019 from Google Play, and collected 1,239,754 reviews for 364 apps\footnote{Since the list of top-200 apps changed during the app crawling, the number of collected apps exceeded 200.} in total.
% We crawled the apps ranked at the top 200 in August 2019
% % \yx{May 2011 $\sim$ August 2019}  
% from Google Play, and collected 1,239,754 reviews for 364 apps in total. 
The collected information for each review instance includes the review text, author name, post date, helpfulness number, and developer's reply. Since the helpfulness numbers for the reviews are updated per day, we remove the repetitive reviews and keep the most recent helpfulness number for each review instance. Consequently, the number of unique review instances is 571,823 and the distribution of helpfulness numbers is shown in Figure~\ref{fig:distribution}. We check the distribution of helpfulness numbers by the Shapiro-Wilk test~\cite{shapiro1965analysis}. Shapiro-Wilk test is a typical test of normality in which the null hypothesis is that the input samples come from a normally distributed population. If the p-value computed by the Shapiro-Wilk test is smaller than 0.05, it means that the input distribution is significantly different from a normal distribution. The Shapiro-Wilk test result (p-value $<0.001$) shows that the helpfulness number is normally distributed. For the convenience of model training, we determine one review is helpful if its helpfulness number exceeds specific quantile $q$. During experiments, we set $q=0.5$ for ensuring that the numbers of helpful and unhelpful reviews in the collected corpus are equally distributed. 

\textbf{Dataset for validating review prioritization.} Following the previous work~\cite{DBLP:conf/icse/VillarroelBROP16}, we exploit the Android user reviews dataset made available by Chen \textit{et al.}~\cite{DBLP:conf/wsdm/ChenHLX15}. The dataset publishes user reviews for multiple releases of $\sim$21K apps, and the information for each review contains the review text, author name, posted date, user rating, and the app's release it refers to. Additionally, each app in the dataset is associated with a metadata file describing its basic information such as the ``updated'' optional field that app developers can use to report the changes they made to the different app releases.
% Similar to~\cite{DBLP:conf/icse/VillarroelBROP16}, we build an oracle reporting which user reviews for release $v_i$ have been implemented by the developers in the release $v_{i+1}$ (i.e., \textit{high priority} reviews), and which have been ignored/postponed (i.e., \textit{low priority} reviews). 
According to Villarroel \textit{et al.}~\cite{DBLP:conf/icse/VillarroelBROP16}, five apps in the CLAP dataset are considered in the work, as shown in Table~\ref{tab:dataset}. The five selected app releases received a total of 11,659 user reviews. Besides, the selected CLAP dataset has no overlapping with the dataset for helpfulness prediction.

% to be labeled as ``implemented'' or ``not implemented'' in $v_{i+1}$. We randomly selected from each app a statistical 

\begin{table}[h]
% 	\small
% 	\captionsetup{aboveskip=-1pt}
	\caption{CLAP dataset for validating review summarization.}
	\label{tab:dataset}
	\center
	\scalebox{1.0}{%
	\begin{tabular}{l l r r}
		\toprule
		App Name & Category & Version No. & \#Reviews \\
		\midrule
		eBay & Shopping & 2.6.1 & 5,210\\
		Viber & Communication & 4.3.1 & 5,878 \\
		Barebone & {Communication} & 3.1.0 & 166 \\
		Hmbtned & {Casual} & 4.0.0 & 71\\
		Timberiffic & {Fashion} & 1.11 & 269\\
        \bottomrule
	\end{tabular}
}
\end{table}

% \yun{Swift key dataset}

% \yun{We also employed the published IDEA dataset for emerging app issue detection in our previous work~\cite{gao2018online} for validating the review prioritization of \tool. The IDEA dataset contains six apps from different app stores and app categories. All the apps include multiple consecutive app versions, with the average review number ranging from 532 to 6,332.}

% \begin{table}[ht]
% % 	\small
% % 	\captionsetup{aboveskip=-1pt}
% 	\caption{IDEA dataset for validating review prioritization. The first two apps are from Apple's App Store and the other apps are from Google Play Store.}
% 	\label{tab:idea_dataset}
% 	\center
% 	\scalebox{1.0}{%
% 	\begin{tabular}{llrr}
% 		%       \rowcolor{gray!50}
% 		\toprule
% 		App Name & Category  & \#Versions & \#Reviews  \\
% 		\midrule
% % 		\thickhline
% 		NOAA Radar & Weather & 16 &  8,363 \\
% % 		\hline
% % 		Waze & Navigation & App Store & 3,849& 9\\
% % 		\hline
% 		YouTube & Multimedia  & 33 & 37,718 \\
% % 		\hline
% % 		Snapchat & Photo \& Video & App Store & 23,073& 11\\
% % 		\hline
% 		Viber & Communication & 8 & 17,126 \\
% % 		\hline
% 		Clean Master & Tools & 7 & 44,327\\
% % 		\hline
% 	    Ebay & Shopping & 9 & 35,483 \\
% 	   % \hline
% 	    Swiftkey & Productivity & 16 & 21,009 \\
% 	    \bottomrule
% 	   % Waze & Navigation & Google Play &9,601 & 12\\
% 	   % \hline
% % 	    SHAREit & Tools & Google Play & 45,316 & 40\\
% % 		\hline
% 	\end{tabular}
% }
% \end{table}

\subsection{Implementation}

% \noindent \textbf{Experiment Conduction} 
The implementation details of the major phases of \tool, including review helpfulness prediction, sentiment-topic modeling, and multi-factor topic and review ranking, are as below. In the review helpfulness prediction phase, we conduct a 10-fold cross-validation for evaluating the trained classifier on distinguishing helpful and unhelpful reviews. In the sentiment-topic modeling phase, we group app reviews into $K=8$ topics for each sentiment polarity. In the multi-factor ranking phase, we take the top 8 reviews for each topic based on the ranking scores for evaluation. Specifically, for the multi-factor topic ranking, we experimentally set $\omega_f$ for the topic's aspects as 0.15, 0.2, 0.35, and 0.3 for the four grading aspects, i.e., topic proportion, topic sentiment, average rating, and freshness, respectively. For the review ranking, we experimentally define $\omega_f$ for the review's aspects as 0.2, 0.1, 0.1, 0.05, 0.05, 0.2, and 0.3 for the aspects including review rating, freshness, negative polarity, neutral polarity, positive polarity, review length, and topic, respectively. 
% We define the default weight of reviews as $\omega_r = [0.2, 0.1, 0.1, 0.05, 0.05, 0.2, 0.3]$. \yx{our reasons}
% The number of words in a review can simply reflect the amount of information contained in the review. The more words a review contains, the more information it contains generally, which is also more helpful for developers to continue to improve their apps.
% And we select top 8 reviews for each topic according to their score computed by the multi-factor ranking (Seen in Section~\ref{sec:ranking}). Finally, we have maximum of 64 reviews in 8 topics to evaluate our approach.
 
In evaluation, we run each result 10 times and computed the average for comparison. We asked three industrial developers who have more than three years of software development experience to manually check the consistency between the prioritized reviews and changelogs, and also the informativeness. Each participant was paid 50\$ for completing the evaluation.
% spent an average of 4 hours for the evaluation.

\begin{table}[htbp]
  \centering
  \caption{Examples of manual evaluation. The columns ``Infor'' and ``Hit'' represents whether the review is informative or semantically consistent with any changelog, respectively.}
  	\scalebox{0.7}{%
    \begin{tabular}{|clcc|}
    \hline
    \multicolumn{4}{|c|}{Changelog of the eBay app}  \\
    \hline
    1     & Search refinement locking now available worldwide &       &  \\
    2     & Fixed bug where seller feedback would not load &       &  \\
    \hline
    \hline
    No. & \multicolumn{1}{c}{Review} & Infor & Hit \\
    \hline
    1     & Wont let me leave feedback & \cmark & \xmark \\
    2     & When I click on view seller's other items, nothing happens. & \cmark & \xmark \\
    3     & It's still can't saved search refinement.  & \cmark   & \cmark \\
    4     & Excellent service and great value for money. & \xmark    & \xmark \\
    5     & Feedback issues & \xmark    & \xmark \\
    \hline
    \end{tabular}%
    }
  \label{tab:manual}%
\end{table}%

We use the eBay app to explain our manual evaluation criteria. Table \ref{tab:manual} presents the changelogs of the app in our benchmark dataset (top) and five review examples (bottom). The criterion for determining 
reviews' informativeness depends on whether the reviews describe issues in detail or provide helpful suggestions for developers. For example, the ``\textit{feedback issues}'' mentioned in review 5 does not contain any useful information for developers, and is labeled as ``non-informative''. The criterion for determining whether reviews hit any changelogs is based on whether the reviews are semantically relevant to the changelogs. For the review 1 in Table \ref{tab:manual}, although the review is related to ``\textit{feedback}'', it describes about ``\textit{leave feedback}'' instead of loading feedback as the second changelog. So review 1 is labeled as inconsistent with the changelogs.

% whether a review is informative is whether it describes the issue in detail or offers helpful suggestions. For example, The "feedback issues" mentioned in the review 5 is not detailed, so the developers can't understand where the issues are. The criterion for determining whether a review matches any changelogs is whether it is semantically relevant rather than keyword relevance. For example, review 1 means buyers can't leave feedback, not that they can't view seller's feedback, therefore it's informative but not match.

% \noindent \textbf{Default Weight Setting} 

\subsection{Evaluation Metrics}
We evaluate the review prioritization results following the previous work~\cite{gao2018online,DBLP:conf/icse/VillarroelBROP16}. The metrics include $Precision$, $Recall$, and $F1\text{-}Score$.

% The evaluation metrics employed in our experiment are divide into two parts based on their features: topic metrics and review metrics. The topic metrics include $Precision$, $Recall$, and $F1\text{-}Score$ that are defined as:

\begin{equation}
    Precision = \frac{\#(G \cap T)}{\#(T)}
\end{equation}
\begin{equation}
    Recall = \frac{\#(G \cap T)}{\#(G)}
\end{equation}
\begin{equation}
    F1\text{-}Score = \frac{2 \times Precision \times Recall}{Precision + Recall}
\end{equation}
\noindent where $T$ and $G$ denote the prioritized topics and changelogs, respectively.

% To validate the performance of review prioritization, 
We also involve the metric $infor\text{-}score$ for measuring the informativeness of the prioritized reviews, defined as:
% and $Hit\text{-}rate$ for evaluating the quality of ranking results defined as:

\begin{equation}
    infor\text{-}score = \frac{\#  informative\;reviews}{\#  prioritized\;reviews},
\end{equation}

% \begin{equation}
%     Hit\text{-}rate = \frac{\#  hitting\;reviews}{\#  prioritized\;reviews},
% \end{equation}

\noindent where $\# prioritized\;reviews$ and $\# informative\;reviews$ indicate the numbers of prioritized reviews and informative reviews among the prioritized reviews, respectively.
% And $\# hitting\;reviews$ represents the reviews that match the key terms in the changelogs.

% Recall and the well-known Normalized Discounted Cumulative Gain (NDCG)~\cite{DBLP:journals/cj/Levene11} for evaluating the quality of ranking results following Chen et al.'s work~\cite{chen2014ar}.

% \begin{equation}
%     NDCG@k=\frac{DCG@k}{IDCG@k},
% \end{equation}

% \noindent where $NDCG@k\in[0,1]$, and higher scores indicate greater agreement between the predicted rank order and the rank order in the groundtruth.

\subsection{Baseline Approaches}
There exist many studies on prioritizing reviews for facilitating release planning. Not all the studies are comparable since some of them involve external knowledge such as source code or GitHub issues, and some require manual annotations for processing. In this paper, we aim to review summarization without manual labeling or external source. To select the baseline approaches for comparison, we examine the related work from several aspects: With/without manually-annotated data (abbreviated as MA data), accessibility of MA data, and with/without external knowledge, reproducibility of source code. We search the related work published in the recent seven years (i.e., 2014$\sim$2021) from Google Scholar\footnote{\url{https://scholar.google.com/}}. To ensure the quality of the papers, we exclude the papers with citations of fewer than five. Table~\ref{tab:literature} lists our examination results. As can be seen in Table~\ref{tab:literature}, all the related approaches require manually-annotated data, which laterally reflects one advantage of \tool, i.e., no manual labor is involved. By removing the prior approaches with external knowledge or unavailability of MA data, we choose AR-Miner and IDEA as baseline approaches.
% \yx{need we explain why we don't compare SURF?}

% We choose  due to the accessibility of MA data and code. Besides, we involve SURF

% SURF as one baseline due to the accessibility of code, and involve CLAP as another baseline due to the similar evaluation dataset as \tool.

% The studies which aim at ranking the most relevant user reviews for GitHub issues or grouping reviews based on manually-labeled data are excluded from the comparison. We compare the effectiveness of \tool with SURF~\cite{di2016would} and AR-Miner~\cite{chen2014ar}.

% To select the baseline approaches for comparison, we examine the prior studies from several aspects: Whether or not involve manually-annotated data, whether the manually-annotated data are available, the reproducibility of source code, and comparability with our approach. The two authors searched the related papers published in the recent five years (i.e., 2014$\sim$2019) from Google Scholar\footnote{\url{https://scholar.google.com/}}.
% We determine the top SE venues according to the core ranking provided by Computing Research \& Education\footnote{\url{http://www.core.edu.au}}. 

\begin{table*}[t]
	\caption{Existing approaches for user review summarization to facilitate app release planning. The term ``manually-annotated data'' is abbreviated as MA data for convenience.}
	\label{tab:literature}
	\center
	\scalebox{1.0}{%
	\begin{tabular}{l c c c c}
	    \toprule
		Approach & \begin{tabular}{@{}c@{}}With MA \\ Data\end{tabular}  & \begin{tabular}{@{}c@{}}Accessibility \\ of MA data\end{tabular}  & \begin{tabular}{@{}c@{}}No External \\ Knowledge\end{tabular}  &  \begin{tabular}{@{}c@{}}Accessibility \\ of Code\end{tabular}\\
		\midrule
		AR-Miner~\cite{chen2014ar} & \cmark & \cmark & \cmark & \xmark \\
		CLAP~\cite{DBLP:conf/icse/VillarroelBROP16,DBLP:journals/tse/ScalabrinoBRPO19} & \cmark & \xmark & \cmark & \xmark \\
% 		PAID~\cite{DBLP:conf/issre/GaoWHZZL15} & Yes & Yes & Yes & No  \\
		CRISTAL~\cite{DBLP:journals/jss/PalombaVBOPPL18} & \cmark & \xmark & \xmark & \xmark\\
% 		IDEA~\cite{gao2018online} & \cmark & \cmark & \cmark & \cmark \\
		Noei \textit{et al.}~\cite{DBLP:journals/ese/NoeiZWZ19} & \cmark & \xmark & \xmark & \xmark \\ 
		SURF~\cite{di2016would} & \cmark & \xmark & \cmark & \cmark\\
% 		IDEA~\cite{gao2018online} & Yes & Yes & Yes & No \\
		Noei \textit{et al.}~\cite{noei2019too} & \cmark & \xmark & \cmark & \xmark \\
		IDEA~\cite{gao2018online} & \cmark & \cmark & \cmark & \cmark\\
		\bottomrule
	\end{tabular}
}
\end{table*}

% \textbf{CLAP}~\cite{DBLP:conf/icse/VillarroelBROP16,DBLP:journals/tse/ScalabrinoBRPO19} is an approach which first categorizes user reviews into three categories, including suggestion for new feature, bug report, and other (such as non-informative reviews), then clusters the related reviews into a single request, and finally recommends which review cluster developers should satisfy in the next release. Both review categorization and cluster recommendation steps demand manually-annotated data for model training. Due to the unavailability of the annotated data and similar benchmark dataset, we directly utilize the performance reported in the original paper. 
% Besides, CLAP is claimed to be more effective than AR-Miner~\cite{chen2014ar}. 

\textbf{AR-Miner}~\cite{chen2014ar} is a typical framework for mining informative app reviews based on informative review extraction and topic modeling, where the prediction for the informativeness of app reviews requires training on manually-labeled data. We adopt the well-trained model for extracting informative reviews of the benchmark dataset, and then conduct review ranking. 

\textbf{IDEA}~\cite{gao2018online} is one of the state-of-the-art online emerging app issue detection approaches. IDEA adapts an online topic modeling approach to track the changes in topics along with app versions, and identifies the abnormal topics as emerging app issues. IDEA automatically labels each topic with the most semantically representative reviews. To ensure a fair completion, we restrict IDEA to predict the topics of the current app version without considering historical app versions.

% \textbf{SURF}~\cite{di2016would,DBLP:conf/icse/SorboPAVC17} is the state-of-the-art approach for condensing the useful information in user feedback. SURF contains a two-level classification model which classifies reviews according to user intentions\footnote{User intention categories include information giving, information seeking, feature request, problem discovery, and other.} (i.e., the first level) and review topics\footnote{The review topics are divided into 12 categories: app, GUI, contents, pricing, feature or functionality, improvement, updates/versions, resources, security, download, model, and company.} (i.e., the second level). Based on the classifier, a summary of ranked user feedback is automatically produced. Although the manually-annotated data are not available, SURF provides an encapsulated tool for automatically processing new reviews.

%% file: sections/experiment.tex
\section{Experimental Results}\label{sec:exper}

In this section, we illustrate the experiment results of \tool by comparing with IDEA~\cite{gao2018online},
and another competing approach, AR-Miner~\cite{chen2014ar}, to assess its capability in prioritizing user reviews. Our experiments are aimed at answering the following research questions:

%  and DIVER~\cite{gao2019diver}

\begin{enumerate}[label=\bfseries RQ\arabic*:,leftmargin=.5in]

    \item What is the impact of different classifiers on the performance of review helpfulness prediction? Which features are more important for differentiating \textit{helpful} reviews from \textit{unhelpful} ones?
    
     \item What is the performance of \tool in app review prioritization compared with the baselines? 

    \item What is the impact of the unhelpful reviews filtering process on the model performance?

    \item What is the impact of different numbers of topics on the performance of \tool?

\end{enumerate}

\subsection{RQ1: Performance of Review Helpfulness Prediction}

% \subsubsection{Motivations for RQ1}
% In review helpfulness prediction phase, we characterize a review instance using the 20 features that we extract. These features are then used to learn a model to distinguish the helpfulness of a user review. 
In RQ1, we explore the efficacy of different classifiers on the performance of review helpfulness prediction. Besides, we study the important features by explicitly considering the contributions of reviews from five dimensions, including stylistics, readability, lexicon, sentiment, and content.
% \yx{I want to explain but I don't know how to compute them}

In this study, we use SVM (Support Vector Machine)~\cite{DBLP:journals/kais/WuKQGYMMNLYZSHS08} as the default classifier to construct the model. We also use Random Forest~\cite{mitchell1997machine} and EMNB (Expectation Maximization Naive Bayes)~\cite{DBLP:journals/ml/NigamMTM00} as the underlying classifiers for our baselines. The prediction results are illustrated in Table~\ref{tab:comparison_help}. 
From the table, we can observe that SVM can attain better overall performance in predicting review helpfulness than RF and EMNB. RF has higher precision than SVM (86.6 v.s. 85.5) but is worse than SVM in recall and F1-score. Besides, SVM and RF outperform EMNB concerning all the metrics. As shown at the bottom of Table~\ref{tab:comparison_help}, we can find without considering any dimension of features reduces the classification performance. Specifically, the stylistics features are the most important for training the classifier regarding the F1-score metric, indicating that helpful reviews tend to present distinguishable text lengths and word lengths compared with unhelpful reviews. Besides, the readability and sentiment features also benefit the classification, which is reasonable. For example, review texts with better readability explain that the reviews are more carefully written and more likely to be helpful.

\begin{table}[h]
	\caption{Review helpfulness prediction results. The bottom of the table presents the ablation study results, where the terms such as ``-Stylistics'' indicate the SVM-based classifier without considering the stylistics features.}
	\label{tab:comparison_help}
	\center
	\scalebox{1.0}{%
	\begin{tabular}{lccc}
		\toprule
		\textbf{Approach}& \textbf{Precision}& \textbf{Recall} & \textbf{F1-score} \\
		\midrule
		EMNB & 72.8&63.9 &68.0 \\
		RandomForest &\textbf{86.6} & 77.9& 82.0\\
		SVM & 85.5& \textbf{78.8}&\textbf{82.2} \\
		\midrule
		\midrule
		\multicolumn{4}{c}{\textbf{SVM-Based}}\\
		\midrule
		-Stylistics & 85.3 & 67.6&75.4 \\
		-Readability  & 83.5& 73.6&78.2 \\
		-Lexicon & 83.8& 76.5& 80.0\\
		-Sentiment & 84.3& 73.8& 78.7\\
		-Content & 82.0& 78.8& 80.4 \\
		\bottomrule
	\end{tabular}
}
\end{table}

\begin{tcolorbox}[width=\linewidth,boxrule=0pt,top=1pt, bottom=1pt, left=1pt,right=1pt, colback=gray!20,colframe=gray!20]
\textbf{Answer to RQ1:} In summary, the SVM-based classifier is effective in review helpfulness prediction. All the features utilized for classification are useful for distinguishing helpful and unhelpful reviews, among which the stylistics features are the most important. 
\end{tcolorbox}

\subsection{RQ2: Comparison with the Baselines}

%no Hit-rate
\begin{table}[ht!]
	\caption{Comparison results with baseline approaches on the CLAP dataset. Bold fonts indicate the best results.}
	\label{tab:baselines}
	\center
	\scalebox{0.8}{%
	\begin{tabular}{clrrrr}
		\toprule
		\begin{tabular}[x]{@{}c@{}}\textbf{App}\\\textbf{Name}\end{tabular} & \textbf{Approach} & \textbf{$infor\text{-}score$} & \textbf{Precision} & \textbf{Recall} & \textbf{F1-Score} \\
		\midrule
		\multirow{2}{*}{eBay} 
        & AR-Miner & 77.81 & \textbf{81.25} & \textbf{100.00}& \textbf{89.50} \\
        & IDEA & 62.46 & 71.25 & \textbf{100.00}& 82.14  \\
		& \tool & \textbf{95.16} & 73.75 & \textbf{100.00} & 84.58  \\
		\midrule
		\multirow{2}{*}{Viber}
        & AR-Miner & 71.41 & 35.00&  44.00 &  37.61 \\
        & IDEA & 70.00 & 48.75 & 42.00& 44.04 \\
		& \tool & \textbf{97.81} & \textbf{66.25} & \textbf{60.00} & \textbf{62.29}  \\
        \midrule
        \multirow{2}{*}{Barebone}
        & AR-Miner & 64.21 & \textbf{62.50}& 53.00&56.05\\
        & IDEA & 55.79 & 40.00 & 46.00 & 41.61 \\
		& \tool & \textbf{80.15} & 61.25 & \textbf{70.00} & \textbf{64.66}  \\
		\midrule
		\multirow{2}{*}{Hmbtned} 
        & AR-Miner  & 41.20 & 89.82 & \textbf{75.00}  & 81.61 \\
        & IDEA & 38.30 & 87.50 & 67.50 & 75.45 \\
		& \tool & \textbf{81.82} & \textbf{96.90} & \textbf{75.00} & \textbf{84.45}  \\
		\midrule
		\multirow{2}{*}{Timeriffic} 
        & AR-Miner & \textbf{78.63} &63.75 & 75.00 & 68.29  \\
        & IDEA & 48.78 & 40.00 & 62.50 & 48.14 \\
        & \tool & 73.26 & \textbf{68.75} & \textbf{86.25} & \textbf{76.13} \\
		\midrule
		\midrule
		\multirow{2}{*}{Average} 
        & AR-Miner & 66.65 &66.46 & 69.40 & 66.61 \\
        & IDEA & 55.07 & 57.50 & 63.60 & 58.28 \\
		& \tool & \textbf{85.64} & \textbf{73.38} &  \textbf{78.25}& \textbf{74.42}\\
		\bottomrule
	\end{tabular}
}
\end{table}

% \subsubsection{Motivations for RQ2}
% To evaluate the performance of \tool in review prioritization, we compare it with baseline approaches including IDEA\cite{gao2018online} and AR-Miner\cite{chen2014ar}. During evaluation, we select the top eight reviews of each topic for \tool and baseline approaches.

To evaluate the performance of \tool in review prioritization, we compare it with baseline approaches including IDEA\cite{gao2018online} and AR-Miner\cite{chen2014ar}. To ensure a fair comparison, we select the top eight reviews of each topic for \tool and baseline approaches.

% We first present the comparison results of \tool with baselines on the CLAP\cite{DBLP:conf/icse/VillarroelBROP16,DBLP:journals/tse/ScalabrinoBRPO19}, then we discuss the performance of \tool from two aspects in the following subsections.
% \subsubsection{Answer to RQ2}

Table \ref{tab:baselines} presents the comparison results on the CLAP datasets. We can observe that \tool performs better than AR-Miner and IDEA in review prioritization.
For example, the average results of \tool are 73.38, 78.25, and 74.42 in terms of $Precision$, $Recall$, and $F1\text{-}Score$, respectively, which outperform 10.41\%, 12.75\% and 11.72\% than AR-Miner, respectively. The results demonstrate that \tool can prioritize more topics containing the key terms in the changelogs, and the prioritized topics also reflect more app changelogs. \tool achieves consistently the best ranking performance regarding the $F1\text{-}Score$ metric for all the studied apps except for eBay. For eBay's reviews, AR-Miner presents slightly better performance than \tool, i.e., 89.50 and 84.58, respectively. The lower results of \tool may be attributed to that the eBay's changelog only describes two changes and may not involve all the changes made in practice~\cite{gao2018online}. We also find that the reviews prioritized by \tool for each topic are more semantically coherent than those results output by AR-Miner, as depicted in Table~\ref{tab:topics}. We choose the two topics ``\textit{seller feedback}'' and ``\textit{search refined}'' since they are semantically consistent with the changelogs of eBay. As can be seen in Table~\ref{tab:topics}, only 25.0\%-37.5\% of the reviews provided by AR-Miner are relevant to the corresponding topics; while the reviews prioritized by \tool are more semantically related. Among all the three approaches, IDEA shows the lowest average performance on our benchmark dataset. This may be because that IDEA is specifically designed for online app review analysis, and may require reviews from multiple historical versions for effective review prioritization.
% , which would hinder the generalizability of IDEA for the apps with reviews of only one version available.

Regarding the $infor\text{-}score$ metric, \tool significantly outperforms the baseline models by at least 28.49\% on average, indicating that the reviews prioritized by \tool are more informative. Future research can utilize \tool to filter non-informative reviews for downstream tasks.
% The experimental results suggest that \tool is more effective in app review .
% can automatic summarization of app reviews.
% We can also find that AR-Miner presents a slightly higher $F1-Score$ than \tool for eBay's reviews (89.50 and 84.58, respectively).
% , which means the average of 6.5/8 topics grouped by AR-Miner during 10 times experiments contain the key terms with eBay's changelogs. 
% High $Precision$ may not fully reflect performance, because it can be improved by reducing the number of topics. And we will discuss it in review metrics.
% Specifically, IDEA performs not very good on CLAP dataset. We suspect the reason to be two-fold: (1) 
% (2) 
% \yx{explain why IDEA is not good}

%no Hit-rate
\begin{table*}[ht!]
	\caption{Comparison on the topic generated by AR-Miner and \tool for the eBay app. The topics are related to ``\textit{seller feedback}'' and ``\textit{search refined}'', respectively, each with top eight reviews presented. Fonts with wavy underlines highlight the terms that not semantically related to the topic.}
	\label{tab:topics}
	\center
	\scalebox{0.9}{%
	\begin{tabular}{lll}
		\toprule
		\multirow{2}{*}{\textbf{Approach}} & \multicolumn{1}{c}{\textbf{Topic 1}} & \multicolumn{1}{c}{\textbf{Topic 2}} \\
		& \multicolumn{1}{c}{\textit{seller feedback}} & \multicolumn{1}{c}{\textit{search refined}} \\
        \midrule
        \multirow{8}{*}{\tool} & ... can't view feedback & ... inability to remember search settings ... \\
        & ... won't load any feedback for sellers ... & ... have to refine search ... \\
        & ... can't see feedback sometimes & ... sick ... change the list results of a search ... \\
        & ... can't read feedback ... from seller ... & ... search result STILL defaults to Best Match ... \\
        & ... \uwave{wouldn't load when connected} ... & ... removed search options ...\\
        & ... can't view message inbox ... & ... have to change everytime I search ...\\
        & ... can't write a message to seller ...& ... search results always ... to ``best match'' ...\\
        & ... can't \uwave{view the descriptions or buy} ...& ... doesn't hold search settings \\
        \midrule
        \multirow{8}{*}{AR-Miner} & ... not \uwave{log in to paypal} to pay ...& ... \uwave{Feedback} won't load ...  \\
        &... have to \uwave{refine search} ... & ... inability to remember search settings ...\\
        & ... Won't let me \uwave{search anything} ...& ... hate the search suggestions ...\\
        & ... can't even \uwave{get on to it} & ... \uwave{pictures don't appear} ... \\
        & ... not able to load mine or seller's feedback & ... \uwave{Glitchy and almost inoperable} ...\\
        & ... when \uwave{search} ... it say's network lost ...& ... Samsung Galaxy S3 \uwave{lose functions} ...\\
        & ... latest seller comments, not just mine ... & ... won't allow \uwave{reinstalling the app} ...\\
        & ... can't \uwave{search} anything ... & ... \uwave{does not open} ...\\
		\bottomrule
	\end{tabular}
}
\end{table*}

\begin{tcolorbox}[width=\linewidth,boxrule=0pt,top=1pt, bottom=1pt, left=1pt,right=1pt, colback=gray!20,colframe=gray!20]
\textbf{Answer to RQ2:} In summary, \tool shows significantly better average performance than baseline approaches in review prioritization. Besides, more than 85\% of the reviews prioritized by \tool are informative, outperforming the baselines by at least 28\%.
\end{tcolorbox}

% no Hit-rate
\begin{table}[ht!]
	\caption{Results of \tool with and without the review helpfulness prediction step. Bold fonts indicate the best results.}
	\label{tab:nofiltering}
	\center
	\scalebox{0.75}{%
	\begin{tabular}{clrrrr}
		\toprule
		\begin{tabular}[x]{@{}c@{}}\textbf{App}\\\textbf{Name}\end{tabular} & \textbf{Approach} & \textbf{$infor\text{-}score$} & \textbf{Precision} & \textbf{Recall} & \textbf{F1-Score} \\
		\midrule
		\multirow{2}{*}{eBay} 
        & \toolnf & 78.59 & 71.25 & \textbf{100.00}& 82.82 \\
		& \tool & \textbf{95.16} & \textbf{73.75} & \textbf{100.00} & \textbf{84.58}  \\
		\midrule
		\multirow{2}{*}{Viber}
        & \toolnf & 58.44 & 32.50 &  50.00 &  39.03 \\
		& \tool & \textbf{97.81} & \textbf{66.25} & \textbf{60.00} & \textbf{62.29} \\
        \midrule
        \multirow{2}{*}{Barebone}
        & \toolnf & 68.51 & \textbf{67.50}& 68.00&\textbf{67.25}\\
		& \tool & \textbf{80.15} & 61.25 & \textbf{70.00} & 64.66  \\
		\midrule
		\multirow{2}{*}{Hmbtned} 
        & \toolnf & 43.93 & 89.82 & 72.50  & 79.65 \\
		& \tool & \textbf{81.82} & \textbf{96.90} & \textbf{75.00} & \textbf{84.45}  \\
		\midrule
		\multirow{2}{*}{Timeriffic} 
        & \toolnf & \textbf{74.59} &\textbf{73.75} & 85.00 & \textbf{77.93}  \\
        & \tool & 73.26 & 68.75 & \textbf{86.25} & 76.13 \\
		\midrule
		\midrule
		\multirow{2}{*}{Average} 
        & \toolnf & 64.81 &66.96 & 75.10 & 69.34 \\
		& \tool & \textbf{85.64} & \textbf{73.38} &  \textbf{78.25}& \textbf{74.42} \\
		\bottomrule
	\end{tabular}
}
\end{table}

\subsection{RQ3: Impact of Review Helpfulness Prediction on the Performance of \tool}

% \subsubsection{Motivations for RQ3}
In this section, we study the impact of the review helpfulness prediction process on the performance of \tool. For analysis, we evaluate the performance of \tool without considering the review helpfulness information for filtering, namely \toolnf. The results are depicted in Table \ref{tab:nofiltering}. We can observe that the reviews prioritized by \tool are more informative than those output by \toolnf, with an increased rate at 32.14\% in terms of the $infor\text{-}score$ metric. The advantage of \tool is consistent for the studied apps, except for the Timeriffic app for which \toolnf only shows marginally higher performance than \tool. Besides, the review helpfulness prediction process contributes greatly to the review prioritization performance. For example, \tool achieves $Precision$, $Recall$, and $F1\text{-}Score$ at 73.38\%, 78.25\%, and 74.42\% on average, respectively, outperforming \toolnf by 9.59\%, 4.19\% and 7.33\%, respectively.

\begin{tcolorbox}[width=\linewidth,boxrule=0pt,top=1pt, bottom=1pt, left=1pt,right=1pt, colback=gray!20,colframe=gray!20]
\textbf{Answer to RQ3:} In summary, the review helpfulness prediction process in \tool is beneficial for filtering non-informative reviews and providing more accurate review summaries. 
\end{tcolorbox}

\subsection{RQ4: Impact of Different Topic Numbers on the Performance of \tool}

\begin{figure}
	\centering
		\includegraphics[width=0.42\textwidth]{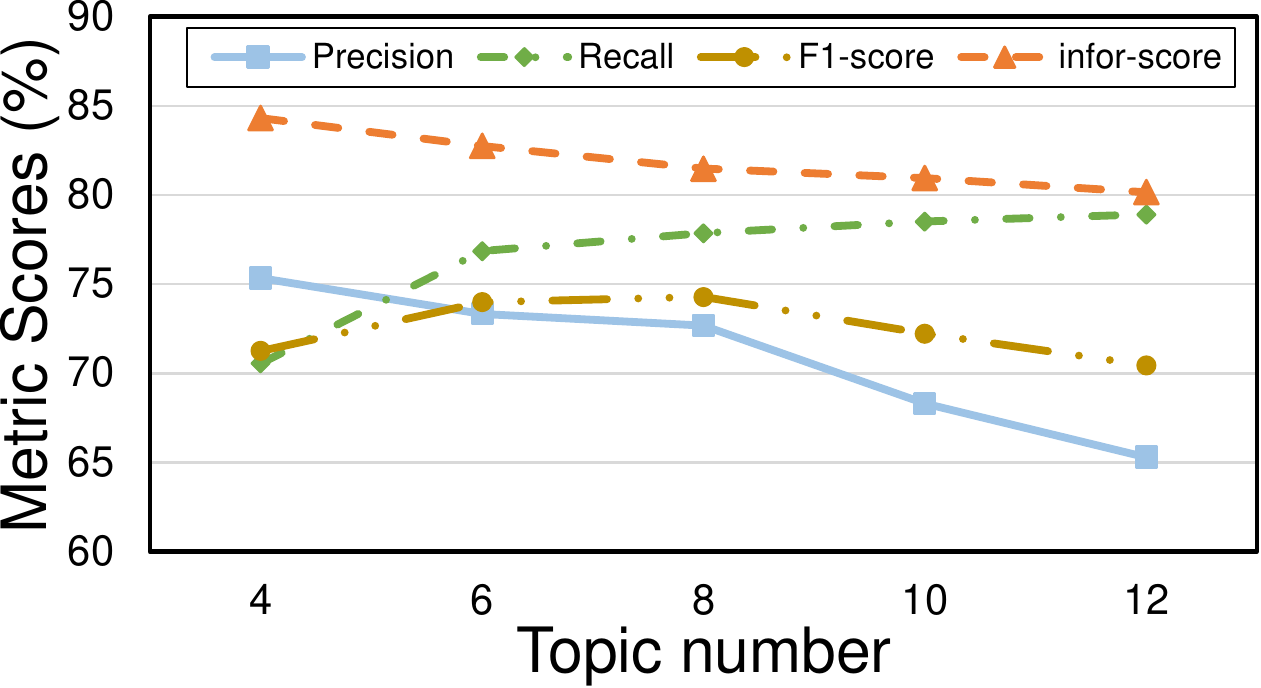}
	\caption{The impact of topic number on the model performance.}
	\label{fig:ntopic}
\end{figure}

% \subsubsection{Motivations for RQ4}
During experimentation, we set the topic number $K=8$ for each sentiment polarity. In this section, we analyze the impact of different topic numbers on the performance of \tool. Figure \ref{fig:ntopic} illustrates the performance changes along with varying topic numbers. We can observe that the values of $Precision$, $F1\text{-}Score$, and $infor\text{-}score$ metrics present a downward trend with the growth of topic number, while the $Recall$ metric shows an increasing trend. The results are reasonable. Larger topic numbers indicate that more reviews are prioritized, and thereby present higher chances to cover more changelogs, leading to an increasing $Recall$ score. Meanwhile, more prioritized reviews would be irrelevant to the changelogs, thus lowering the other metric scores. As can be seen in Figure~\ref{fig:ntopic}, \tool achieves relatively better performance when the topic number is defined as 6 or 8. During experimentation, we set the topic number as 8 according to the $F1\text{-}Score$ metric.

\begin{tcolorbox}[width=\linewidth,boxrule=0pt,top=1pt, bottom=1pt, left=1pt,right=1pt, colback=gray!20,colframe=gray!20]
\textbf{Answer to RQ4:} In summary, \tool generally shows a downward trend with the growth of topic number. The model achieves relatively better performance when the topic number is defined as 6 or 8. According to the F1\text{-}Score metric, we set the topic number as 8.
\end{tcolorbox}

%% file: sections/discussion.tex
\section{Discussion}\label{sec:discussion}

\subsection{Threat and Validity}
There are four major threats to the validity of our study.
\begin{enumerate}
%   \item The diversity and freshness of available datasets. We directly use the publicly released data of CLAP provided by their authors. The data include only 5 apps from Google Play Store. The limited categories and number of studied apps may influence the generalization of the proposed \tool. Besides, the dataset for helpfulness prediction model is up to August 2019. Since most of the previous studies have not released data or the purposes are not consistent with our focus, we will eliminate this threat as soon as more datasets are publicly available.

    \item The diversity and freshness of available datasets. We directly use the publicly released data of CLAP provided by their authors. The data include only 5 apps from Google Play Store. The limited categories and number of studied apps may influence the generalization of the proposed \tool. Besides, the helpfulness prediction model dataset was created in 2019, which seems a bit old. Since the recently published review data~\cite{wu2021identifying,DBLP:conf/re/HenaoFSFV21} do not involve the helpfulness number, we train the helpfulness prediction model based on the old dataset. Moreover, the features of helpful reviews from different periods would be similar, so the freshness of the reviews would not be a great threat. We will conduct more experiments when appropriate datasets get publicly available.

    \item Bias in manual evaluation. For checking the performance of \tool, we invite three industrial developers to evaluate the consistency between the prioritized reviews and changelogs, and also the informativeness. The results of the human evaluation can be impacted by the participants' experience. To mitigate the bias in human evaluation, we ensure that all three different participants evaluated each prioritized review. Besides, all the participants are industrial developers who have more than three years of software development experience.

    \item Evaluation of baseline models. 
    % Replication of baseline models. 
    For comparison, we survey the recent studies on app review analysis, and chose two reproducible baselines AR-Miner~\cite{chen2014ar} and IDEA~\cite{gao2018online}. Since the original papers do not report the results on our benchmark datasets, we evaluate the baselines by carefully replicating\footnote{\url{https://github.com/monsterLee599/AR-Miner}} the algorithms described in the original work of AR-Miner and restricting IDEA to prioritize reviews without considering historical app versions.
    
    % Replication of baseline models. For evaluation, we survey the recent studies on app review analysis, and chose two reproducible baselines AR-Miner~\cite{chen2014ar} and IDEA~\cite{gao2018online} for comparison. Since the original papers do not report the results on our benchmark datasets, we carefully replicate the baselines following the algorithms described in the papers. Besides, we run each result 10 times and computed the average to ensure the fairness.
    
    % Randomness in a single experiment. In our approach, we use an unsupervised model, named BST, for jointly modeling topics and sentiment of user reviews. Because probability distributions influence the topic modeling process, there is no guarantee that the outcomes of each experiment will be the same. To mitigate this risk, we run each result 10 times and computed the average, which can reduce the error of a single experiment.
    
    \item Weights in the multi-factor topic and review ranking. In the multi-factor ranking phase, the weights in Equ. (7) and Equ. (12) for respectively computing rankings scores of topics and reviews can impact the performance of the proposed approach. In this work, we experimentally set the weights for evaluation, indicating that the reported results of \tool may be sub-optimal. In future work, we will build upon heuristic algorithms~\cite{DBLP:journals/tec/HuangLY20} to automatically determine the optimal weights.
    % \yx{Weighting parameters in the multi-factor ranking. In the multi-factor ranking phase, the weight $\omega_f$ for computing topic and review ranking score should be manually defined, which can influence the performance of our approach. In practice, we plan to employ heuristic algorithms\cite{huang2019survey} to determine the optimal weight in the future.}
    
\end{enumerate}

\subsection{Analysis on the Impact of Rating Normalization}

\begin{table}[ht!]
	\caption{Results of \tool without and with rating normalization during review ranking, where \toolnn indicates that the review ranking process does not involve rating normalization. Bold fonts indicate the best results.}
	\label{tab:normalized}
	\center
	\scalebox{0.73}{%
	\begin{tabular}{clrrrr}
		\toprule
		\begin{tabular}[x]{@{}c@{}}\textbf{App}\\\textbf{Name}\end{tabular} & \textbf{Approach} & \textbf{$infor\text{-}score$} & \textbf{Precision} & \textbf{Recall} & \textbf{F1-Score} \\
		\midrule
		\multirow{2}{*}{eBay} 
        & \toolnn & 93.25& \textbf{80.00} & \textbf{100.00}& \textbf{88.20} \\
		& \tool & \textbf{95.16} & 73.75 & \textbf{100.00} & 84.58  \\
		\midrule
		\multirow{2}{*}{Viber}
        & \toolnn & 86.41 & 57.50 &  \textbf{60.00} &  57.96 \\
		& \tool & \textbf{97.81} & \textbf{66.25} & \textbf{60.00} & \textbf{62.29} \\
        \midrule
        \multirow{2}{*}{Barebone}
        & \toolnn & 79.89 & \textbf{61.25}& \textbf{70.00} &\textbf{65.10}\\
		& \tool & \textbf{80.15} & \textbf{61.25} & \textbf{70.00} & 64.66  \\
		\midrule
		\multirow{2}{*}{Hmbtned} 
        & \toolnn  & \textbf{81.82} & 82.83 & \textbf{75.00}  & 78.41 \\
		& \tool & \textbf{81.82} & \textbf{96.90} & \textbf{75.00} & \textbf{84.45}  \\
		\midrule
		\multirow{2}{*}{Timeriffic} 
        & \toolnn & 72.41 & \textbf{72.50} & \textbf{87.50} & \textbf{79.10}  \\
        & \tool & \textbf{78.63} & 68.75 & 86.25 & 76.13 \\
		\midrule
		\midrule
		\multirow{2}{*}{Average} 
        & \toolnn & 83.76 & 70.73 & \textbf{78.50} & 73.77 \\
		& \tool & \textbf{85.64} &\textbf{73.38} &  78.25 & \textbf{74.42} \\
		\bottomrule
	\end{tabular}
}
\end{table}

During ranking reviews in Section III-D, we conduct normalization on the ratings, as shown in Equ. (13). In this section, we analyze the impact of rating normalization on the performance of \tool. The results are illustrated in Table~\ref{tab:normalized}, where \toolnn indicates the review ranking without rating normalization.
The average results of \toolnn are 70.73, 78.50, and 73.77 in terms of $Precision$, $Recall$, and $F1\text{-}Score$ on average, respectively.
According to the experimental results, we can observe that \tool{} performs slightly better than \toolnn in $Precision$, $F1\text{-}Score$ and $infor\text{-}Score$. The results indicate rating normalization during review ranking can help \tool{} to prioritize informative reviews.

\subsection{Analysis on the Positive and Neutral Sentiment Polarity of Reviews}

To explore the effects of positive and neutral sentiment polarity of reviews, we only consider the negative sentiment polarity of reviews (namely \toolon{}), and Table~\ref{tab:only} presents the results. The average results of \toolon are 74.73, 77.25, and 75.00 in terms of $Precision$, $Recall$, and $F1\text{-}Score$, respectively. Compared to \tool{}, we can observe that only consider the negative sentiment polarity has a  slightly effect on the $Precision$, $Recall$, and $F1\text{-}Score$ scores, but has a significant impact on the $infor\text{-}score$. This result indicates that the positive and neutral sentiment polarity of reviews can help \tool{} to prioritize more informative reviews.

\begin{table}[ht!]
	\caption{Results of \tool only considering negative polarity and all the sentiment polarities during review ranking, where \toolnf indicates that the review ranking only involves negative reviews. Bold fonts indicate the best results.}
	\label{tab:only}
	\center
	\scalebox{0.75}{%
	\begin{tabular}{clrrrr}
		\toprule
		\begin{tabular}[x]{@{}c@{}}\textbf{App}\\\textbf{Name}\end{tabular} & \textbf{Approach} & \textbf{$infor\text{-}score$} & \textbf{Precision} & \textbf{Recall} & \textbf{F1-Score} \\
		\midrule
		\multirow{2}{*}{eBay} 
        & \toolon  & 89.84 & \textbf{83.75} & \textbf{100.00}& \textbf{90.72} \\
		& \tool & \textbf{95.16} & 73.75 & \textbf{100.00} & 84.58  \\
		\midrule
		\multirow{2}{*}{Viber}
        & \toolon & 79.69 & 58.75 &  60.00 & 58.25 \\
		& \tool & \textbf{97.81} & \textbf{66.25} & \textbf{60.00} & \textbf{62.29} \\
        \midrule
        \multirow{2}{*}{Barebone}
        & \toolon & 79.29 & \textbf{63.75}& \textbf{70.00}&\textbf{66.25}\\
		& \tool & \textbf{80.15} & 61.25 & \textbf{70.00} & 64.66  \\
		\midrule
		\multirow{2}{*}{Hmbtned} 
        & \toolon & \textbf{81.82} & 92.40 & \textbf{75.00}  & 82.55 \\
		& \tool & \textbf{81.82} & \textbf{96.90} & \textbf{75.00} & \textbf{84.45}  \\
		\midrule
		\multirow{2}{*}{Timeriffic} 
        & \toolon & 71.84 &\textbf{75.00} & 81.25 & \textbf{77.25}  \\
        & \tool & \textbf{73.26} & 68.75 & \textbf{86.25} & 76.13 \\
		\midrule
		\midrule
		\multirow{2}{*}{Average} 
        & \toolon & 80.49 & \textbf{74.73} & 77.25 & \textbf{75.00} \\
		& \tool & \textbf{85.64} & 73.38 &  \textbf{78.25}& 74.42 \\
		\bottomrule
	\end{tabular}
}
\end{table}

\subsection{Analysis on the Number of Selected Reviews}

\begin{figure}
	\centering
		\includegraphics[width=0.42\textwidth]{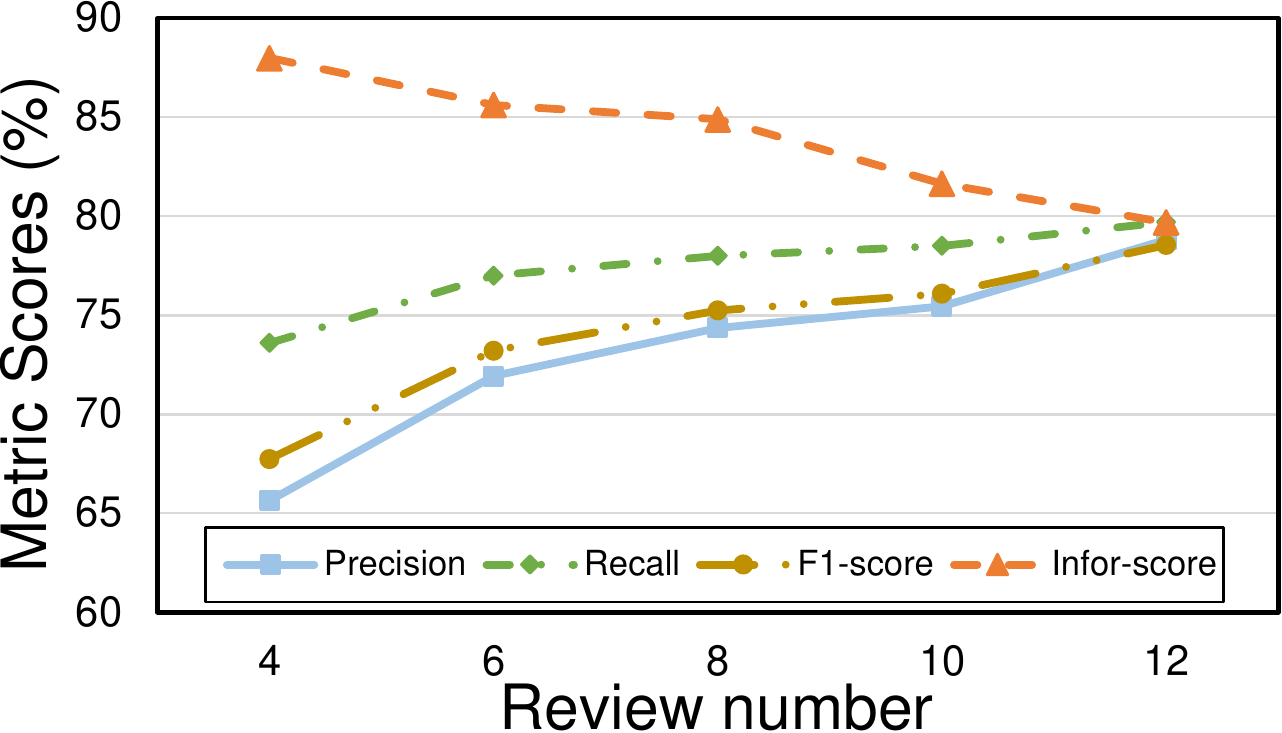}
	\caption{The impact of the number of prioritized reviews on the model performance.}
	\label{fig:nreview}
\end{figure}

In this section, we study the impact of different numbers of prioritized reviews on the performance of \tool. The results are illustrated in Figure~\ref{fig:nreview}.
We can observe that the values of $Precision$, $Recall$, and $F1\text{-}Score$ present an increasing trend along with the growth of the review number, while the $infor\text{-}score$ metric shows a downward trend. The results are reasonable since larger numbers of prioritized reviews tend to cover more changelogs; however, they are also likely to include more non-informative reviews. To balance the informativeness and accuracy of prioritized reviews, we choose the number of prioritized reviews as 8 during experimentation.

% The number of selected reviews can also impact the performance of evaluation. In this section, we study the impact of different numbers of prioritized reviews on the performance of \tool, and Figure~\ref{fig:nreview} illustrates the performance changes along with varying review numbers for each topic.
% We can observe that the values of $Precision$, $Recall$ and $F1\text{-}Score$ present a increasing trend with the growth of review number, while the $infor\text{-}score$ metric shows an downward trend. 
% Generally speaking, a topic with more reviews is more likely to cover more changelogs, improving the $Precision$, $Recall$ and $F1\text{-}Score$ scores. But it will also include some reviews with lower review ranking score, resulting in a lower $infor\text{-}score$. As can be seen in Figure~\ref{fig:nreview}, \tool achieves relatively better performance when using the top 8 or 10 reviews.

%% file: sections/literature.tex
\section{Related Work}\label{sec:literature}
\subsection{App Marketplace Analysis}

The growth of smartphones and mobile applications makes the app marketplace a hotspot for researchers within and outside the software engineering community. Harman \textit{et al.} \cite{DBLP:conf/msr/HarmanJZ12} pointed out that app marketplaces provide a wealth of information in the form of pricing and customer reviews and thus can be treated as a new form of software repository. They also used data mining to analyze apps' technical, customer and business aspects in BlackBerry World. Chia \textit{et al.} \cite{DBLP:conf/www/ChiaYA12} discovered that the ratings used in app marketplaces are not reliable indicators of privacy risks of an app. Minelli \textit{et al.} \cite{DBLP:conf/csmr/MinelliL13} proposed to leverage source code, usage of third-party APIs, historical data, along with data extracted from app marketplace to better comprehend apps. Martin \textit{et al.} \cite{DBLP:conf/sigsoft/MartinSH16} introduced an approach to causal impact analysis to help app developers understand the impact of app releases. They also conduct a comprehensive survey on app marketplace analysis, including review mining.

\subsection{App Review Mining}

User feedback plays an essential role in serving as a major channel between developers and users, reflecting new feature requirements, enhancements in the user interface, and reporting serious app bugs~\cite{dkabrowski2022analysing}. For many years, researchers from academia and industry have explored mining app reviews for assisting different stages of app development and maintenance, such as prioritizing app reviews \cite{chen2014ar,DBLP:conf/issre/GaoWHZZL15,DBLP:conf/issre/ManGLJ16,DBLP:journals/jss/PalombaVBOPPL18,malgaonkar2022prioritizing}, predicting app feature liked/disliked by users \cite{DBLP:conf/kbse/GuK15,DBLP:conf/re/GuzmanM14}, classifying app reviews \cite{DBLP:conf/icse/VillarroelBROP16,di2016would,DBLP:conf/re/MaalejN15}, and identifying emerging app issues~\cite{gao2019diver,gao2018online}.

The booming user reviews inspired researchers to come up with heuristic approaches. Regarding prioritizing app issues, Chen \textit{et al.} \cite{chen2014ar} proposed a computational framework that visualizes the most ``informative'' reviews which are identified by a topic model and an effective review ranking scheme. Gao \textit{et al.} \cite{DBLP:conf/issre/GaoWHZZL15} pointed out that the issues presented in the level of phrase, i.e., a couple of consecutive words, can be more easily understood by developers than in long sentences. Then they designed a framework to track reviews over the release versions of the app and recommend phrase-level issues of an app to its developers. Malgaonkar \textit{et al.} \cite{malgaonkar2022prioritizing} studied recent works on app review prioritization and developed a multi-criteria heuristic model for identifying and prioritizing informative reviews. For predicting app features liked/disliked by users, Gu \textit{et al.} \cite{DBLP:conf/kbse/GuK15} and Guzman \textit{et al.} \cite{DBLP:conf/re/GuzmanM14} proposed to classify reviews into predefined categories and extracts aspects in sentences that include evaluation of aspect using natural language processing techniques.  In order to classify reviews into different categories and prioritize emerging issues. Villarroel \textit{et al.} \cite{DBLP:conf/icse/VillarroelBROP16} proposed a framework to categorize user reviews based on the information they carry out (e.g., bug reporting), cluster together related reviews (e.g., all reviews reporting the same bug), and automatically prioritize the clusters of reviews to be implemented. Gao \textit{et al.} \cite{gao2019diver,gao2018online} proposed an efficient and automated framework to identify emerging app issues based on online review analysis which achieves both high accuracy and real-time identification. Wu \textit{et al.}~\cite{wu2021identifying} created a Chinese dataset from the Chinese Apple App Store and built a regression model to identify key features of app by analyzing app description and positive/negative user reviews. Haering \textit{et al.}~\cite{haering2021automatically} focused on the gap between technically-written bug reports with colloquially-written app reviews, extracting issues from app reviews and matching them to bug reports. 
Henao \textit{et al.}~\cite{henao2021transfer} proposed a framework for mining feature requests and bug reports from tweets and app store reviews via transfer learning.

In recent years, researchers are getting into analyzing the dynamic nature of user reviews.
For example, Gao \textit{et al.}~\cite{gao2018online} automatically capture app issues discussed in user reviews and detect the emerging ones for version modification.
Besides employing user feedback for collecting user opinions, Guzman \textit{et al.}~\cite{DBLP:conf/icse/GuzmanIG17} incorporate app-related twitters to facilitate the software evolution process.
Nayebi \textit{et al.}~\cite{DBLP:conf/esem/NayebiFR17} propose the concept of ``marketability'' for open source mobile apps, and adopt analogical reasoning to guide unsuccessful marketable releases to be transited into successful ones.

% \textcolor{blue}{The sentiment analysis techniques aim to detect the polarity (positive or negative) of sentiment embedded in the text\cite{cambria2017affective,zhang2018deep,yadav2020sentiment}, and make significant progress via graph convolutional networks in recent years~\cite{dai2022learning,liang2022aspect,zhao2022graph}. Given the increased research interest in intelligence app, more and more NLP approaches, e.g., sentiment analysis, were employed in app review mining\cite{panichella2016ardoc,luiz2018feature}. For example, Luiz \textit{et al.}~\cite{luiz2018feature} adopted the sentiment analysis strategy to identify the sentiment associated with each topic, and discuss the importance in app review mining. Yang \textit{et al.}~ \cite{yang2021tour} proposed TOUR, a tool for dynamic topic and sentiment analysis of user reviews for assisting app release. 
% In addition, sentiment analysis can be applied to filter the neutrality~\cite{valdivia2018consensus} reviews and  handle ambivalence~\cite{wang2020multi} reviews.}

% Automatic review summarization
Automatic review summarization is another challenging problem in app review mining because most app reviews are short, noisy, non-informative, and sometimes contain multiple and various topics for different apps \cite{chen2014ar}. Natural language processing approaches have been adopted to tackle this challenge. Previous research papers \cite{noei2019too, di2016would} identified common topics in app reviews by different granularity, such as searching, web browsing, pricing, and resources.
% For example, Noei \textit{et al.} \cite{noei2019too} identified 23 common topics, such as searching and web browsing; while Di Sorbo \textit{et al.} \cite{di2016would} summarized 12 topic clusters, including pricing and resources, etc., which are more general compared to Noei \textit{et al.}'s definition \cite{noei2019too}.
Mudambi \textit{et al.} \cite{DBLP:conf/hicss/MudambiSZ14} found that not all the topics demand developers' deep inspection. Besides, ratings of user reviews are a commonly-used index, but the ratings may be aligned with the review texts. Therefore, accurate prioritization of the topics can be time-saving. Fu \textit{et al.}~\cite{DBLP:conf/kdd/FuLLFHS13} filtered reviews that expressed inconsistent sentiment with their ratings and then summarized the remaining topics. Iacob \textit{et al.} \cite{DBLP:conf/msr/IacobH13} utilised Latent Dirichlet Allocation \cite{DBLP:conf/nips/BleiNJ01} and linguistic rules to generate summary for new feature requests.
Araújo \textit{et al.} \cite{de2021re} proposed a BERT-based language model to automatically extract software requirements from app reviews.

\subsection{Sentiment Analysis}
The sentiment analysis techniques aim to detect the polarity (e.g., positive, neutral, or negative) of sentiment implied by texts~\cite{DBLP:journals/expert/Cambria16,DBLP:journals/air/YadavV20}.
In recent years, many studies apply deep learning models, including reinforcement learning~\cite{DBLP:journals/inffus/PengMPLC21}, emotional recurrent unit~\cite{DBLP:journals/ijon/LiSJC22}, and graph convolutional networks~\cite{DBLP:journals/ijdsa/DaiHNC22,DBLP:journals/kbs/LiangSGCX22}, for sentiment analysis~\cite{DBLP:journals/widm/ZhangWL18,DBLP:conf/emnlp/PoriaCG15,SenticNet7}. 
There exists other techniques~\cite{DBLP:journals/inffus/ValdiviaLCH18,DBLP:journals/ijufks/WangHC20} proposed for sentiment analysis. 
For example, Valdivia \textit{et al.} \cite{DBLP:journals/inffus/ValdiviaLCH18,DBLP:conf/fuzzIEEE/ValdiviaLH17} proposed weighted aggregation models for detecting and filtering neutral texts.
Wang \textit{et al.} \cite{DBLP:journals/ijufks/WangHC20}
proposed a multi-level fine-scaled approach to handle ambivalence in the text.
% Nevertheless, deep learning models are also challenged by some complex situations in sentiment analysis~\cite{DBLP:journals/air/YadavV20}, e.g., the ambiguity in review~\cite{DBLP:journals/ijufks/WangHC20}. 
% Besides, sentiment analysis can be employed for more complex and challenging situations, e.g., detecting and filtering neutrality~\cite{DBLP:journals/inffus/ValdiviaLCH18} and handling ambivalence~\cite{DBLP:journals/ijufks/WangHC20}.
% Sentiment analysis has proven useful for facilitating app review mining\cite{DBLP:conf/sigsoft/PanichellaSGVCG16,DBLP:conf/www/LuizVAMSCGR18}, and 
We will consider the issue of ambivalence~\cite{DBLP:journals/ijufks/WangHC20} in our scenario in the future.